\documentclass[aps,prd,groupedaddress,preprint,eqsecnum,nofootinbib]{revtex4}
\usepackage{graphicx,epsf,amssymb,amsbsy,amsfonts,amssymb,amsmath}

\usepackage{slashbox}

\newif\ifdraft
\drafttrue
\newif\ifpreprint
\preprinttrue

\def\be{\begin{equation}}
\def\ee{\end{equation}}
\def\bea{\begin{eqnarray}}
\def\eea{\end{eqnarray}}
\def\beal{\begin{equation}\begin{aligned}}
\def\eeal{\end{aligned}\end{equation}}
\def\nn{\nonumber}

\def\spa#1.#2{\left\langle#1\,#2\right\rangle}
\def\spb#1.#2{\left[#1\,#2\right]}

\def\eqn#1{eq.~\eqref{#1}}

\begin{document}
\hfuzz 20pt

\ifpreprint
UUITP-24/17
\fi

\title{Ambitwistor formulations of $R^2$ gravity and $(DF)^2$ gauge theories}

\author{Thales~Azevedo and Oluf~Tang~Engelund}

\affiliation{
Department of Physics and Astronomy, Uppsala University, 75108 Uppsala, Sweden 
}

\email{{thales.azevedo,  oluf.engelund}@physics.uu.se}

\date{October, 2017}

\begin{abstract}
We consider $D$-dimensional amplitudes in $R^2$ gravities (conformal gravity in $D=4$) and in the recently introduced $(DF)^2$ gauge theory, from the perspective of the CHY formulae and ambitwistor string theory.  These theories are related through the BCJ double-copy construction, and the $(DF)^2$ gauge theory obeys color-kinematics duality.  We work out the worldsheet details of these theories and show that they admit a formulation as integrals on the support of the scattering equations, or alternatively, as ambitwistor string theories.  For gravity, this generalizes the work done by Berkovits and Witten on conformal gravity to $D$ dimensions. The ambitwistor is also interpreted as a $D$-dimensional generalization of Witten's twistor string (SYM + conformal supergravity). As part of our ambitwistor investigation, we discover another $(DF)^2$ gauge theory containing a photon that couples to Einstein gravity. This theory can provide an alternative KLT description of Einstein gravity compared to the usual Yang-Mills squared.
\end{abstract}

\maketitle

\tableofcontents

\section{Introduction}

In a fascinating paper \cite{Cachazo:2013hca}, Cachazo, He and Yuan constructed a way to write the $n$-point amplitudes for Yang-Mills and for gravity in $D$ dimensions. They wrote the amplitudes as $n$-dimensional integrals on the support of the so-called scattering equations. Later on, the formalism was shown to be well-suited to describe other theories as well, such as bi-adjoint scalars \cite{Cachazo:2013iea}, Dirac-Born-Infeld and the non-linear sigma model \cite{Cachazo:2014xea}. These compact formulae were subsequently shown to also arise from ambitwistor strings \cite{Mason:2013sva,Casali:2015vta}.

\begin{table}
\begin{center}
\begin{tabular}{|c|c|c|c|c|}
\hline
\backslashbox{\begin{tabular}{l}Left\\Action\end{tabular}}{\begin{tabular}{r}Right\\Action\end{tabular}}&\begin{tabular}{c}Current\\Algebra\end{tabular}&\begin{tabular}{c}Single\\Fermion\end{tabular}&\begin{tabular}{c}Two\\Fermions\end{tabular}&None\\
\hline
\begin{tabular}{c}Current\\Algebra\end{tabular}&\begin{tabular}{c}Bi-adjoint\\scalar\end{tabular}&Yang-Mills&\begin{tabular}{c}Non-linear\\sigma model\end{tabular}&$(DF)^2$\\
\hline
\begin{tabular}{c}Single\\Fermion\end{tabular}&Yang-Mills&\begin{tabular}{c}Einstein\\Gravity\end{tabular}&Born-Infeld&\begin{tabular}{c}Conformal\\Gravity\end{tabular}\\
\hline
\begin{tabular}{c}Two\\Fermions\end{tabular}&\begin{tabular}{c}Non-linear\\sigma model\end{tabular}&Born-Infeld&Galileon&$(DF)^2$-photon\\
\hline
None&$(DF)^2$&\begin{tabular}{c}Conformal\\Gravity\end{tabular}&$(DF)^2$-photon&$\rm(Weyl)^3$\\
\hline
\end{tabular}
\end{center}
\caption{The matrix of ambitwistor string actions with the new row/column (None).\label{teorimatrix}}
\end{table}

In this paper we will add three extra theories to the list of those that admit a simple CHY-type formulation. The first theory is the $(DF)^2$ theory constructed in \cite{conformal}. This theory is related to conformal gravity \cite{Fradkin:1985am} via the KLT relations \cite{Kawai:1985xq}. We compute its lower point amplitudes and subsequently find an $n$-point generalization, that possesses the correct factorization channels. The CHY formulation makes the absence of all $\varepsilon_i\cdot\varepsilon_j$-terms in the amplitudes manifest, a property that is otherwise obscure from a Feynman diagram representation.

The second theory we consider is conformal gravity itself\footnote{or more accurately a $D$-dimensional $R^2$ theory which in $D=4$ becomes conformal gravity. Throughout the paper we will use these terms interchangeably. There are other types of $R^2$ gravity but this is the only one of interest to us}. For this theory we also propose a CHY formulation for the $n$-point amplitude and show that it factorizes correctly. The formula beautifully generalizes the one by Berkovits and Witten for conformal gravity \cite{Berkovits:2004jj}, to which it reduces when considering the MHV sector in $D=4$.

%These theories might be regarded as somewhat ``sick'', given the presence of modes with a wrong-sign propagator which render the theories non-unitary. However, there are many reasons why they are worth studying. For instance, conformal supergravities are candidates for a fundamental theory of gravity, i.e. a theory describing physics at Planck distances \cite{Fradkin:1985am}. Moreover, they share some features with their Poincar\'e cousins, such as [insert something about U(1) anomaly paper \cite{Carrasco:2013ypa} here], and can be related to ordinary gravity in asymptotically (anti-) de Sitter space \cite{Maldacena:2011mk}. The $\alpha^\prime \to \infty$ limit of the heterotic string should also be related to some kind of conformal gravity.
%As for the $(DF)^2$ gauge theory, besides being a double-copy constituent of conformal gravity, it is of interest for the ambitwistor string community since it helps clarify some aspects of the theory, as we show in this paper.

Finally, we show that these theories can be given a straightforward interpretation in terms of ambitwistor strings. In our investigation of the corresponding ambitwistor string theories, we find a third theory which can be given a simple CHY formulation. This theory consists of a photon field governed by a $(DF)^2$ term and coupled to Einstein gravity. With these theories in hand, we can expand the usual matrix of possible ambitwistor theories with a new row/column. The new matrix of ambitwistor theories is shown in table \ref{teorimatrix}, with different choices of ambitwistor actions and the resulting theories coming from these actions. Note that the (Weyl)${}^3$-theory is just the usual bosonic ambitwistor string, corresponding to the choice (None, None).

At tree level, the theories we analyze can also be interpreted as sectors of previously considered ambi\-twistor models. For example, the conformal gravity given by the (Single Fermion, None)-choice is a sector of the heterotic ambitwistor string given by (Single Fermion, Current Algebra), in the same sense that Berkovits--Witten is a sector of Witten's twistor string \cite{Witten:2003nn} . In fact, the same is true for any pair of theories of the form \{(X, None); (X, Current Algebra)\}.
Nonetheless, it is remarkable that the ambitwistor approach allows us to truncate the larger models and consider those sectors themselves as stand-alone theories,
and the applicability of this is exemplified by the fact that the theories considered in this paper had not been discussed before in the context of ambitwistor strings.

%For these theories we will propose $n$-point CHY formulae for the amplitudes and show how these theories can be given a straightforward ambitwistor interpretation. This allows us to expand the usual matrix of possible ambitwistor theories with a new row/column. In doing so we discover a third theory which consists of Einstein gravity coupled to a photon governed by a $(DF)^2$ term. The expanded matrix of ambitwistor theories is shown in figure \ref{teorimatrix} with different choices of ambitwistor actions and the resulting theories of these actions.

%perform multiple checks and construct corresponding ambitwistor string theories. As part of our investigation, we find a theory of $(DF)^2$ photons coupled to Einstein gravity that is naturally related to the other two theories.
%We will propose an $n$-point formula for the amplitudes of both theories, check that they match the Feynman diagram results for small $n$ and show that they factorize correctly. These amplitudes will also prove to easily admit an ambitwistor string interpretation.

% After determining the $n$-point CHY formulation for the amplitudes of the two theories, we will construct

One should note that the theories studied in this paper are un-physical, due to the presence of modes with a wrong-sign propagator which render the theories non-unitary. However, they are interesting to study because of their relationships with well-known, physical theories. Conformal supergravity can be related to Einstein gravity in asymptotically (anti-) de Sitter space \cite{Maldacena:2011mk} and its $U(1)$ anomaly can be used to study the similar anomaly in Poincar\'e supergravity \cite{Carrasco:2013ypa}. % Conformal supergravities are also candidates for a fundamental theory of gravity, $i.e.$ a theory describing physics at Planck distances 
 Furthermore, the $\alpha^\prime \to \infty$ limit of the heterotic string should also be related to some kind of conformal gravity. The theory with a $(DF)^2$ photon coupled to Einstein gravity, which we mentioned above and will describe further on in section \ref{ambitwist sec}, is also related to a physical theory. By taking a specific limit, it is possible (at tree level) to relate the amplitudes of the photons to graviton amplitudes from pure Einstein gravity. This provides an alternative route for generating (tree-level) gravity amplitudes through the double copy by merging the $(DF)^2$ theory of Johansson and Nohle with the non-linear sigma model. As for the $(DF)^2$ itself, apart from being a piece in the double-copy constructions, it is of interest for the ambitwistor string community since it helps clarify some aspects of the theory, as we show in this paper.

The paper is structured as follows. We will begin by describing some basic properties of gluon amplitudes, the $(DF)^2$ theory and the similarities between the amplitudes of this theory and those of Yang-Mills (section \ref{DFtheory}). Then we will review the scattering equations and the CHY-formulation of amplitudes, as well as some functions that will prove useful later (section \ref{scat eq sec}). We will then argue that the theories in question give rise to amplitudes that are extremely simple when written in the CHY-formulation (section \ref{amp sec}). Subsequently we show how these simple formulae can arise from ambitwistor theories (section \ref{ambitwist sec}). Finally we sum up our results in the conclusions.

\section{\boldmath{$(DF)^2$ Theory}}\label{DFtheory}

The $(DF)^2$ theory created by Johansson and Nohle \cite{conformal} will play an essential role in this paper so in this section the theory will briefly be described. We should perhaps note that Lagrangians with similar operators have previously been studied for phenomenological reasons in \cite{Simmons:1989zs,Simmons:1990dh,Cho:1993eu,Duff:1991ad,Dreiner:1991xi,Dixon:1993xd} and because the operators arise as corrections in the $\alpha'$ expansion of bosonic open string theory \cite{Barreiro:2012aw,Barreiro:2013dpa,Boels:2016xhc}. It is however the specific theory introduced in \cite{conformal} that interests us as it satisfies color-kinematics duality and gives conformal gravity through the double copy. In general the amplitudes of this theory have many features similar or identical to the beautiful features of Yang-Mills amplitudes. For this reason it will be useful to review some of the basic properties of Yang-Mills amplitudes.

For starters the tree-level amplitudes of gluons in Yang-Mills theory can be written as a sum over single-trace color factors and corresponding color-ordered amplitudes:

\begin{align}
{\cal A}_n^{\rm tree}&=g^{n-2}\sum_{{\rm perm} (2,3,\ldots n)} {\rm Tr}(T^{a_1} T^{a_2} T^{a_3} \cdots T^{a_n}) \, A(1,2,3, \ldots,n).\label{c-o amp}
\end{align}

Using the Kleiss-Kuijf relations \cite{Kleiss:1988ne}, this can be re-expressed as a sum over strings of structure constants:

\begin{align}
{\cal A}_n^{\rm tree}&=(ig)^{n-2}\sum_{{\rm perm} (2,3,\ldots n-1)} f^{a_1a_2b_1}f^{b_1a_3b_2}\cdots f^{b_{n-3}a_{n-1}a_n} \, A(1,2,3, \ldots,n),
\end{align}

\noindent where the color-ordered amplitudes are the same as in \eqref{c-o amp}. This is known as the DDM basis \cite{DelDuca:1999iql,DelDuca:1999rs}, and it is the form that the amplitudes from the ambitwistor string naturally appear in.

The gluon amplitudes of Yang-Mills are also known to satisfy the color-kinematics duality \cite{Bern:2008qj} (see also \cite{Sondergaard:2009za}), which works as follows. Consider an $n$-point amplitude written in the form:

\begin{align}
\mathcal{A}_n=&{}(ig)^{n-2}\sum_{i\ \in\ \textrm{cubic graphs}}\frac{n_i c_i}{D_i},\label{amp for c-k}
\end{align}

\noindent where the $c_i$'s are products of structure constants, the $n_i$'s are kinematic numerators and the $D_i$'s are products of propagators. There is a certain ambiguity in how the numerators are chosen because the $c_i$'s are dependent on each other due to the Jacobi relations. However the color-kinematics duality tells us that it is possible to chose the numerators in such a way that they satisfy relations identical to the Jacobi relations for the corresponding color factors.

For the color-ordered amplitudes the duality leads to the BCJ relations \cite{Bern:2008qj} (proven from a string theory perspective in \cite{BjerrumBohr:2009rd} and from a field theory perspective in \cite{Feng:2010my} using the BCFW recursion relations \cite{Britto:2004ap,Britto:2005fq}).

\begin{align}
0=p_1\cdot p_2 A(1,2,3, \ldots,n)+\sum_{i=3}^{n-1} (p_1\cdot p_2 + p_2\cdot p_3+\cdots +p_2\cdot p_i) A(1,3,\cdots, i,2,i+1, \cdots,n).\label{BCJ amp}
\end{align}

Writing the amplitudes in a form satisfying color-kinematics duality has the advantage that it makes the relationship between Yang-Mills and Einstein gravity straightforward. If the numerators satisfy the duality, one simply replaces the color factors, $c_i$, by another copy of the numerators, $n_i$, in order to arrive at the amplitudes for gravity. This is known as the double copy and is equivalent to the KLT relations:

\begin{align}
{\cal M}_n^{\rm EG}&= A^{\rm YM} \cdot S \cdot A^{\rm YM}\,,\label{KLT EG}
\end{align}

\noindent where the color-ordered gauge-theory amplitudes have been packaged into column/row vectors of $(n-3)!$ size, and the matrix $S$ is the (field theory) KLT kernel.

Schematically we can write this as:

\begin{align}
{\rm EG} &= {\rm YM} \otimes {\rm YM}\,.
\end{align}

These are the properties of Yang-Mills theory that will be relevant for our discussion of the $(DF)^2$ theory which we will now turn to. The Lagrangian of this theory is given by:

\begin{align}
{\cal L}_{(DF)^2}=&{}   \frac{1}{6}(D_{\mu} F^{a\, \mu \nu})(D^{\rho} F^{a}_{\phantom{a}\, \rho \nu}) + \frac{1}{3} (D^{\rho} F^{a\, \mu \nu})(D_{\mu} F^{a}_{\phantom{a}\, \rho \nu})   + \frac{1}{2}g \,  C^{\alpha ab}  \varphi^{ \alpha}   F_{\mu \nu}^a F^{b\, \mu \nu }  \\
&+\frac{1}{2}(D_{\mu} \varphi^{\alpha})^2+ \frac{1}{3!} g \, d^{\alpha \beta \gamma}   \varphi^{ \alpha}  \varphi^{ \beta} \varphi^{ \gamma}\nn.
\end{align}
where the field strength and the covariant derivatives are defined as
\begin{align}
%F^3 &=& f^{abc} F^{a\, \mu}_{\nu}  F^{b\, \nu}_{\rho} F^{c\, \rho}_{\mu}  \nn \\
F_{\mu \nu}^a &= \partial_{\mu} A_{\nu}^a-\partial_{\nu} A_{\mu}^a + g  f^{abc} A_{\mu}^b A_{\nu}^c, \nn \\
D_{\rho} F_{\mu \nu}^a &= \partial_{\rho} F_{\mu \nu}^a +  g  f^{abc} A_{\rho}^b F_{\mu \nu}^c  , \label{fieldDef}\\
D_{\mu} \varphi^\alpha  &= \partial_{\mu} \varphi^\alpha -  i g  (T^{a})^{\alpha \beta} A_{\mu}^a \varphi^\beta  \nn.
\end{align}

The scalar $\varphi^\alpha$ transforms in a real representation of the gauge group, with generator $(T^a)^{\alpha \beta}$. Some of the interactions are parametrized by symmetric Clebsch-Gordan coefficients $C^{\alpha ab}=C^{\alpha ba}$ and totally symmetric $d^{\alpha \beta \gamma}$ constants, which are only implicitly defined through the two relations

\begin{align}
&C^{\alpha ab}C^{\alpha cd} = f^{ace}f^{edb}+ f^{ade}f^{ecb}\,, \nn \\
&C^{\alpha ab}d^{\alpha \beta \gamma}= (T^a)^{\beta \alpha} (T^b)^{\alpha \gamma}+ C^{\beta ac} C^{\gamma cb} + (a \leftrightarrow b)\,.
\label{colorConstr}
\end{align}

%These color-relations are necessary for the $(DF)^2$ theory to obey color-kinematics duality; likewise the relative coefficients in front of the individual operators in the Lagrangian are fixed by color-kinematics duality. 

From \eqn{colorConstr}, and together with the Lie algebra relations that trivially follow from infinitesimal group transformations
\begin{align}
&(T^{a})^{\alpha \gamma}(T^{b})^{\gamma \beta}-(T^{b})^{\alpha \gamma}(T^{a})^{\gamma \beta}= i f^{abc} (T^{c})^{\alpha \beta}\,,   \label{1stID} \\
&f^{bae}C^{\alpha ec}+f^{cae}C^{\alpha be}=i(T^{a})^{\alpha \beta}C^{\beta bc}\,, \label{2ndID}  \\
&(T^{a})^{\alpha \delta}d^{\delta \beta \gamma}+(T^{a})^{\beta \delta}d^{\alpha \delta \gamma}+(T^{a})^{\gamma \delta}d^{\alpha \beta \delta}=0  \label{3rdID} \,,
\end{align}

\noindent we have a sufficient number of relations to reduce any tree-level Feynman diagram with external adjoint particles (and possibly internal scalars) to a sum over strings of $f^{abc}$ structure constants, or equivalently, a sum over single-trace factors ${\rm Tr}(T^{a_1} \cdots T^{a_n})$. So the gluonic amplitudes for this theory can also be expressed as in \eqref{c-o amp}. Furthermore, the color-ordered amplitudes will obey the Kleiss-Kuijf relations by virtue of the fact that the trees can alternatively be expressed in terms of only $f^{abc}$'s. Hence it is also possible to express the amplitudes of the $(DF)^2$ theory in the DDM basis as well.

Of course there are significant differences between Yang-Mills and the $(DF)^2$ theory. For instance, the $(DF)^2$ theory will have $1/p^4$ poles since the kinetic term has four derivatives, and in four dimensions the all-plus and single-minus amplitudes are non-vanishing $A(\pm + + \ldots+)\neq0$. The latter implies that the theory does not admit a supersymmetric generalization, which can also be seen from the presence of the $F^3$ term in the Lagrangian; this operator is well-known to be incompatible with supersymmetry.

Besides the gluon and scalar states, the $(DF)^2$ contain gluon ghost states (i.e. the linearized equations of motion for $A^\mu$ has additional solutions) which have the wrong-sign propagator. According to standard field-theory arguments this suggest that the $(DF)^2$ theory violates unitarity; however, this will not be important in the current context. As formal objects the tree amplitudes are well defined, and it is not surprising that such ghost states are present given the close relationship between $(DF)^2$ and conformal gravity. The only caveat is that we need to be careful with how the gluon amplitudes are defined. The external gluon states are taken on the usual plane-wave form $\varepsilon^\mu e^{i p \cdot x}$, and for the LSZ prescription we are amputating the Feynman diagrams by isolating the residue of the $1/p^4$ poles of the external legs.

Some examples of four-gluon amplitudes in $D=4$ are
\begin{align}
A^{(DF)^2}(1^-,2^-,3^+,4^+)&=2i u \frac{\spa{1}.{2}^2}{\spa{3}.{4}^2}\,,\nn \\
A^{(DF)^2}(1^-,2^+,3^-,4^+)&=2i  u \frac{\spa{1}.{3}^2}{\spa{2}.{4}^2}\,, \nn \\
A^{(DF)^2}(1^+,2^+,3^+,4^+)&=2i u \frac{\spb{1}.{2} \spb{3}.{4}}{\spa{1}.{2}\spa{3}.{4}}\,,\nn \\
A^{(DF)^2}(1^-,2^+,3^+,4^+)&=2i  \spb{2}.{4}^2  \frac{\spa{1}.{2} \spb{2}.{3}}{\spb{1}.{2}\spa{2}.{3}}\,.
\end{align}

Notice how some of the color-ordered amplitudes have $1/p^4$ poles and one of them has a $u$ pole which is not possible in Yang-Mills for this particular ordering. These amplitudes however still satisfy the BCJ amplitudes relations \eqref{BCJ amp} and it is possible to write the amplitudes in such a form that they satisfy color-kinematics duality (the relations \eqref{colorConstr} are necessary for the theory to satisfy the duality, and demanding that the theory satisfy the duality was part of how the color relations were found in \cite{conformal}). Notice that the denominators, $D_i$, in \eqref{amp for c-k} will still be the same as they were in Yang-Mills theory, even though this theory contains double propagators. The extra poles will simply be absorbed into the numerator factors.

As shown in \cite{conformal} it is possible to get conformal gravity by using the double copy between the $(DF)^2$ and ordinary Yang-Mills. Schematically, we write this as
\begin{align}
{\rm CG} &= (DF)^2 \otimes {\rm YM}\,.
\end{align}
For the supersymmetric generalizations (${\cal N}=1,2,4$ in $D=4$ notation) we get conformal supergravity from the double copy 
\begin{align}
{\rm CSG} &= (DF)^2 \otimes {\rm SYM}\,,
\end{align}
where all the supersymmetry belongs to the SYM theory. At tree level and for adjoint external particles, we can write the double copy in terms of the KLT formula,
\begin{align}
{\cal M}_n^{\rm C(S)G}&= A^{(DF)^2} \cdot S \cdot A^{\rm (S)YM}\,.\label{KLT CSG}
\end{align}

As an example consider the following four-point MHV amplitude in conformal gravity
\begin{align}
M^{\rm CG}(1^{--},2^{--},3^{++},4^{++}) &=A^{(DF)^2}(1^-,2^-,3^+,4^+) \Big( -i \frac{st}{u}\Big) A^{\rm YM}(1^-,2^-,3^+,4^+)= i  \frac{\spa{1}.{2}^4 \spb{3}.{4}^4}{s^2}\,.
\end{align}

One can of course do the double copy where both numerators come from the $(DF)^2$ theory. As will hopefully become clear in section \ref{ambitwist sec}, the resulting theory will be the $(\rm Weyl)^3$ theory that arises from the bosonic ambitwistor string \cite{Mason:2013sva}.

\section{The Scattering Equations and the CHY Formula}\label{scat eq sec}

It is our goal to express the amplitudes of the theory described in section \ref{DFtheory} in the CHY formulation. In this section we will therefore review some basics about the CHY formulation as well as some functions that will prove useful when considering the $( DF)^2$ theory.

The amplitudes of several quite different theories can be written in the following form in $D$ dimensions:
%As noticed in \cite{Cachazo:2013hca} the amplitudes of Yang-Mills in any dimension can be written as an integral on the support of the scattering equations:

\begin{align}
\mathcal{A}_n=&{}ig^{n-2}\int \frac{d^n\sigma}{\mathrm{vol[SL}(2,\mathbb{C})]}\prod_i\phantom{}'\delta\left(\sum_{j\neq i}\frac{p_i\cdot p_j}{\sigma_{ij}}\right)%\sum_{\beta\in S_n/Z_n}\frac{\mathrm{Tr}\left(T^{a_{\beta(1)}}T^{a_{\beta(2)}}\cdots T^{a_{\beta(n)}}\right)}{\sigma_{\beta(1)\beta(2)}\sigma_{\beta(2)\beta(3)}\cdots \sigma_{\beta(n)\beta(1)}}\mathrm{Pf}'M_n
I_LI_R\label{CHY generel}
\end{align}

Here the prime on the product sign means that three of the delta function are left out:

\begin{align}
\prod_i\phantom{}'\delta\left(\sum_{j\neq i}\frac{p_i\cdot p_j}{\sigma_{ij}}\right)\equiv&{}\sigma_{kl}\sigma_{lm}\sigma_{mk}\prod_{i\neq k,l,m}\delta\left(\sum_{j\neq i}\frac{p_i\cdot p_j}{\sigma_{ij}}\right).
\end{align}

This is necessary as the scattering equations are $\mathrm{SL}(2,\mathbb{C})$ invariant. The three factors of $\sigma$ in the above expression ensures invariance under permutations. Similarly the factor of $\mathrm{vol[SL}(2,\mathbb{C})]$ in the denominator is also necessary in order not to integrate over infinitely many identical terms. It indicates that three of the integration variables will have to be fixed. The remaining part of the integrand is divided into two parts: a left integrand and a right integrand. When we turn towards the ambitwistor string theories, these two parts of the integrand will correspond to different parts of the string action.

In order to get Yang-Mills amplitudes, one can make the following choices for the left and right integrand:

\begin{align}
I_L=&{}\sum_{\beta\in S_n/Z_n}\frac{\mathrm{Tr}\left(T^{a_{\beta(1)}}T^{a_{\beta(2)}}\cdots T^{a_{\beta(n)}}\right)}{\sigma_{\beta(1)\beta(2)}\sigma_{\beta(2)\beta(3)}\cdots \sigma_{\beta(n)\beta(1)}},&I_R=&{}\mathrm{Pf}'M_n.\label{YM CHY}
\end{align}

The dependence on the polarization vectors in the amplitude comes from the $2n\times2n$ antisymmetric matrix called $M_n$. This matrix can be written in the following form:

\begin{align}
M_n=&{}\left(\begin{array}{cc}
M_A&-M_C^T\\
M_C&M_B
\end{array}
\right),\label{MatrixM_n}
\end{align}
\noindent where the different submatrices are defined as:
\begin{align}
M_{A,n}^{i,j}&=\left\{\begin{array}{cc}\frac{p_i\cdot p_j}{\sigma_{ij}}&\text{for }i\neq j\\
0&\text{for }i=j\end{array}\right.,&M_{B,n}^{i,j}&=\left\{\begin{array}{cc}\frac{\varepsilon_i\cdot \varepsilon_j}{\sigma_{ij}}&\text{for }i\neq j\\
0&\text{for }i=j\end{array}\right.\label{SubMatricesM_n}\\
M_{C,n}^{i,j}&=\left\{\begin{array}{cc}\frac{\varepsilon_i\cdot p_j}{\sigma_{ij}}&\text{for }i\neq j\\
-\sum_{k\neq i}\frac{\varepsilon_i\cdot p_k}{\sigma_{ik}}&\text{for }i=j\end{array}\right..\nonumber
\end{align}

The Pfaffian of this matrix vanishes so the object appearing in the CHY formula is the reduced Pfaffian which is defined by removing rows and columns number $k$ and $l$, then computing the Pfaffian of this smaller matrix and finally multiplying by $(-1)^{k+l}/\sigma_{kl}$. The choice of $k$ and $l$ is arbitrary.

If one instead is interested in the amplitudes of Einstein gravity, one can choose both the left and the right integrand to be given by reduced Pfaffians:

\begin{align}
I_L=&{}\mathrm{Pf}'M_n,&I_R=&{}\mathrm{Pf}'M_n.\label{Einstein Gravity}
\end{align}

If on the other hand, one chooses both the left and the right integrand to be given by a color trace over a Parke-Taylor factor:

\begin{align}
I_L=&{}\sum_{\beta\in S_n/Z_n}\frac{\mathrm{Tr}\left(T^{a_{\beta(1)}}T^{a_{\beta(2)}}\cdots T^{a_{\beta(n)}}\right)}{\sigma_{\beta(1)\beta(2)}\sigma_{\beta(2)\beta(3)}\cdots \sigma_{\beta(n)\beta(1)}},&I_R=&{}\sum_{\beta\in S_n/Z_n}\frac{\mathrm{Tr}\left(T^{a_{\beta(1)}}T^{a_{\beta(2)}}\cdots T^{a_{\beta(n)}}\right)}{\sigma_{\beta(1)\beta(2)}\sigma_{\beta(2)\beta(3)}\cdots \sigma_{\beta(n)\beta(1)}},
\end{align}
\noindent one will end up with the amplitudes of a bi-adjoint scalar.

\subsection{Some useful building blocks}

In order to write the amplitudes for the $(DF)^2$ theory from section \ref{DFtheory} in the CHY form, it is necessary to use some additional building blocks, besides the ones that Yang-Mills and gravity amplitudes are constructed from. These building blocks must contain an additional factor of momentum squared as compared to the reduced Pfaffian used for Yang-Mills amplitudes. This can easily be seen by inspecting the Lagrangian: the term with three gluons also contains three derivatives (as opposed to one for Yang Mills), the term with four gluons contains two derivatives (as opposed to none for Yang Mills) etc. Fortunately such factors have already been discussed in the literature \cite{Lam:2016tlk,He:2016iqi}. They can be written in terms of the following functions:

\begin{align}
w_{(i_1i_2\cdots i_k)}=&{}\frac{\frac{1}{2}\mathrm{tr}\left(f_{i_1}f_{i_2}\cdots f_{i_k}\right)}{\sigma_{i_1i_2}\sigma_{i_2i_3}\cdots \sigma_{i_ki_1}},\label{Psi more indices}
\end{align}
where the trace is over Lorentz indices and the $f$'s are linearized field strengths:

\begin{align}
f_{i}^{\mu\nu}=&{}p_{i}^\mu\varepsilon_{i}^\nu-p_{i}^\nu\varepsilon_{i}^\mu.
\end{align}

One also needs to introduce the following special case:

\begin{align}
w_{(i)}=&{}-\sum_{j\neq i}\frac{\varepsilon_i\cdot p_j}{\sigma_{ij}}\label{Psi one index}.
\end{align}

A useful feature of these functions is that they are gauge-invariant. Equation \eqref{Psi more indices} is manifestly gauge-invariant because the linearized field strengths are while equation \eqref{Psi one index} is gauge-invariant on the support of the scattering equations. It can however be a good idea to rewrite \eqref{Psi one index} in order to make M\"obius invariance manifest in the formula for the amplitude. Therefore we employ momentum conservation to re-express the function as:

\begin{align}
w_{(i)}=&{}\sum_{j\neq i}\frac{\varepsilon_i\cdot p_j\sigma_{jr}}{\sigma_{ri}\sigma_{ij}}.
\end{align}

Here $r$ is simply some external leg which is different from $i$. This is a better way of writing the function because $\sigma_i$ then appears twice in the denominator just like in the other functions in \eqref{Psi more indices}, making it easier to construct manifestly M\"obius invariant quantities.

From the above elements we construct the following permutationally invariant functions to be used when constructing the $n$-pt. amplitudes:

\begin{align}
W_{i_1i_2\cdots i_k}=&{}\sum_{\beta\in S_n}\frac{1}{i_1i_2\cdots i_r}w_{(\beta(1)\cdots \beta(i_1))}w_{(\beta(i_1+1)\cdots \beta(i_1+i_2))}\cdots w_{(\beta(i_1+i_2\cdots+i_{k-1}+1)\cdots \beta(n))}\label{P functions}
\end{align}

Here the $i$'s are chosen to satisfy:

\begin{align}
&i_1\leq i_2\cdots\leq i_k,&i_1+i_2\cdots+i_k&=n.
\end{align}

These functions have exactly the right number of momenta: $W_{111}$, $W_{12}$ and $W_3$ all contain three momentum vectors while $W_{1111}$, $W_{112}$, $W_{22}$, $W_{13}$ and $W_4$ all contain four momentum vectors. This exactly matches the counting mentioned above. We thus expect that the right integrand will consist of these functions in place of the reduced Pfaffian while the left integrand will remain the color trace over a Parke-Taylor factor just like in Yang-Mills. Indeed this expectation will turn out to be correct.

One should notice that the functions defined in equation \eqref{P functions} are not all independent. They can be combined to give the Pfaffian of the matrix $M_n$ which as mentioned before is zero:

\begin{align}
M_n=&{}\sum_{1\leq i_1\leq i_2\cdots \leq i_k\leq n}(-1)^{n-k}W_{i_1i_2\cdots i_k}.\label{vanishing Pfaffian}
\end{align}

As a consequence of this one gets that:

\begin{align}
0=&{}W_{111}-W_{12}+W_3,\nonumber\\
0=&{}W_{1111}-W_{112}+W_{13}+W_{22}-W_4,\label{relations among Ws}\\
0=&{}W_{11111}-W_{1112}+W_{113}+W_{122}-W_{14}-W_{23}+W_5,\nonumber\\
0=&{}W_{111111}-W_{11112}+W_{1113}+W_{1122}-W_{114}-W_{123}-W_{222}+W_{15}+W_{24}+W_{33}-W_6.\nonumber
\end{align}

Because of these relations there can be different ways of expressing the amplitudes. We will try to write the amplitudes in a way that makes the generalization to $n$-point amplitudes as straigthforward as possible.

\section{The Amplitudes}\label{amp sec}

Having described the CHY formalism as well as some functions that will prove useful, we can now turn our attention to the amplitudes of the $(DF)^2$ theory described in section \ref{DFtheory}. We have computed the amplitudes up to 6 points using standard Feynman rules and then subsequently determined which of the previously described functions matched them. The expressions in the CHY formalism were evaluated using the tools developed in \cite{Baadsgaard:2015voa,Bjerrum-Bohr:2016juj} (how to apply these tools to double poles has also been dealt with in \cite{Zhou:2017mfj}). We arrive at the following results for the amplitudes:

\begin{align}
\mathcal{A}_3^{(DF)^2}=&{}{-}4ig\int\!\!\! \frac{d^3\sigma}{\mathrm{vol[SL}(2,\mathbb{C})]}\prod_i\phantom{}'\delta\left(\sum_{j\neq i}\frac{p_i\cdot p_j}{\sigma_{ij}}\right)\sum_{\beta\in S_3/Z_3}\frac{\mathrm{Tr}\left(T^{a_{\beta(1)}}T^{a_{\beta(2)}}T^{a_{\beta(3)}}\right)}{\sigma_{\beta(1)\beta(2)}\sigma_{\beta(2)\beta(3)}\sigma_{\beta(3)\beta(1)}}W_{111},\label{Amp3}\\
\mathcal{A}_4^{(DF)^2}=&{}{-}4ig^2\int\!\!\! \frac{d^4\sigma}{\mathrm{vol[SL}(2,\mathbb{C})]}\prod_i\phantom{}'\delta\left(\sum_{j\neq i}\frac{p_i\cdot p_j}{\sigma_{ij}}\right)\sum_{\beta\in S_4/Z_4}\frac{\mathrm{Tr}\left(T^{a_{\beta(1)}}T^{a_{\beta(2)}}\cdots T^{a_{\beta(4)}}\right)}{\sigma_{\beta(1)\beta(2)}\sigma_{\beta(2)\beta(3)}\cdots \sigma_{\beta(4)\beta(1)}}W_{1111},\\
\mathcal{A}_5^{(DF)^2}=&{}{-}4ig^3\int\!\!\! \frac{d^5\sigma}{\mathrm{vol[SL}(2,\mathbb{C})]}\prod_i\phantom{}'\delta\left(\sum_{j\neq i}\frac{p_i\cdot p_j}{\sigma_{ij}}\right)\sum_{\beta\in S_5/Z_5}\frac{\mathrm{Tr}\left(T^{a_{\beta(1)}}T^{a_{\beta(2)}}\cdots T^{a_{\beta(5)}}\right)}{\sigma_{\beta(1)\beta(2)}\sigma_{\beta(2)\beta(3)}\cdots \sigma_{\beta(5)\beta(1)}}W_{11111},\\
\mathcal{A}_6^{(DF)^2}=&{}{-}4ig^4\int\!\!\! \frac{d^6\sigma}{\mathrm{vol[SL}(2,\mathbb{C})]}\prod_i\phantom{}'\delta\left(\sum_{j\neq i}\frac{p_i\cdot p_j}{\sigma_{ij}}\right)\sum_{\beta\in S_6/Z_6}\frac{\mathrm{Tr}\left(T^{a_{\beta(1)}}T^{a_{\beta(2)}}\cdots T^{a_{\beta(6)}}\right)}{\sigma_{\beta(1)\beta(2)}\sigma_{\beta(2)\beta(3)}\cdots \sigma_{\beta(6)\beta(1)}}W_{111111}.
\end{align}

%All the amplitudes seem to follow a simple pattern except $\mathcal{A}_3^{(DF)^2}$. However notice that since the special 3-point kinematics set all products of momenta equal to zero, the different 3-point functions are almost identical:

As mentioned, equations \eqref{relations among Ws} allow us to write the amplitudes in different ways. At 3 points there is furthermore the additional property that all products of momenta are zero because of the special 3-point kinematics. This means that the functions at 3-point become proportional to each other, $W_3\propto W_{12}\propto W_{111}$. There are therefore several ways of representing the amplitudes. The reason for the choices above is that they expose a rather simple pattern which is easy to generalize to $n$ points.

Based on the amplitudes above, we propose the following expression for the $n$-point amplitude:

%\begin{align}
%W_3\bigg|_{p_i\cdot p_j=0}=2W_{111},\\
%W_{12}\bigg|_{p_i\cdot p_j=0}=3W_{111}.
%\end{align}

%This can be used to write \eqref{Amp3} in a manner similar to the others. It is then natural that the $n$-point amplitude will take the following form:

\begin{align}
\mathcal{A}_n^{ (DF)^2}\!\!=&{}{-}4ig^{n-2}\int \frac{d^n\sigma}{\mathrm{vol[SL}(2,\mathbb{C})]}\prod_i\phantom{}'\delta\left(\sum_{j\neq i}\frac{p_i\cdot p_j}{\sigma_{ij}}\right)\!\!\sum_{\beta\in S_n/\sigma_n}\!\!\frac{\mathrm{Tr}\left(T^{a_{\beta(1)}}T^{a_{\beta(2)}}\cdots T^{a_{\beta(n)}}\right)}{\sigma_{\beta(1)\beta(2)}\sigma_{\beta(2)\beta(3)}\cdots \sigma_{\beta(n)\beta(1)}}W_{\underbrace{11\cdots1}_n}\label{DF^2 CHY},
\end{align}

This equation is somewhat similar to the formula for the Yang-Mills amplitudes, only with the reduced Pfaffian, $\mathrm{Pf}'M_n$, replaced by the function $4W_{11\cdots1}$. A curious property of this formula is that it contains no $\varepsilon_i\cdot\varepsilon_j$-terms. This has interesting consequences upon dimensional reduction. Consider the case where we go from $D$ dimensions to $d$ dimensions. The $D$-dimensional gluon then splits into a $d$-dimensional gluon and $D-d$ scalars. However the lack of any $\varepsilon_i\cdot\varepsilon_j$-terms in the amplitudes tells us that the new scalars decouple. This property is not manifest in the Feynman rules and only appears after many different terms cancel each other.

In order to support the claim that \eqref{DF^2 CHY} is in fact the correct $n$-point generalization, we are going to check that it has the correct factorization channels. Since we only have a formula for the scattering of $n$ gluon fields and none with the scalars in the theory as external states, we are going to focus on how the amplitude factorizes when a gluon goes on-shell. These are in any case the easiest factorization channels to determine since they provide a double pole when $q^2\to0$ as opposed to the scalars which only give a single pole.

%In support of this expression for the $n$-point amplitude, we will show that it has the correct factorization channels.

%Using this identity it seems that the natural n-point generalization is given  by simply replacing $\mathrm{Pf}'M_n$ in \eqref{YM CHY} with $4P_{11\cdots1}$:

%We will now show that this guess has the correct factorization channels.

\subsection{Factorization}\label{fact sub}

In order to check the factorization channels of \eqref{DF^2 CHY}, the external momenta are divived into two groups:

\begin{align}
L&=\{1,\cdots, n_L\},&R&=\{n_L+1,\cdots, n\}.
\end{align}

We then consider the case where the sum of the momenta in each group goes on-shell:

\begin{align}
\sum_{i=1}^{n_L}p_i&\equiv q_R,\\
\sum_{i=n_L+1}^np_i&\equiv q_L=-q_R,\\
q_R^2&\to0.
\end{align}

If \eqref{DF^2 CHY} is the correct $n$-point generalization, the formula should develop a $q_R^{-4}$-pole and the residue of this pole be the product of two lower-point amplitudes of the same form. %The scalars in the theory will also give rise to poles. They will only be $q_R^{-2}$-poles so it will not be possible to detect when a gluon with the same momentum goes on-shell. However the color factors coming from the scalars are different from the gluons and in fact a scalar propagator can give rise to a pole even if the particles in 
This will turn out to indeed be the case as can be demonstrated by considering different pieces of the formula individually.

As shown in \cite{Cachazo:2013iea}, the trick to study a factorization channel like the one above is to redefine the integration variables:

\begin{align}
\sigma_i&=\frac{s}{u_i},&\text{for }i&\in L,\\
\sigma_i&=\frac{v_i}{s},&\text{for }i&\in R.
\end{align}

The variables $u_1$, $u_2$ and $v_n$ will be fixed in order to remove the $\mathrm{SL}(2,\mathbb{C})$ symmetry from the amplitude expression. In addition to this, the variable $v_{n-1}$ will be consider to be fixed in exchange for treating $s$ as an integration variable. This means that now four $u$, $v$ variables are fixed. However one would expect there to be six (three for each amplitude). The last two of the fixed integration variables will be the ones corresponding to the new states arising from letting $q_R^2$ go on-shell, and in the calculations to come the quantities will factorize into pieces that will look exactly as expected if the $u$ and $v$ variables corresponding to the new on-shell states have been set to zero.

The $s$ integration will be responsible for the pole. When $q_R^2$ goes to zero, the variable will begin to behave like

\begin{align}
s^2&\sim\frac{q_R^2}{\sum_{i\in R}\frac{v_n-v_i}{v_n}\sum_{j\in L}\frac{2p_i\cdot p_j}{u_jv_i}}\label{s^2 behave}.
\end{align}

The order of the pole (or whether there is one) then depends on how many factors of $s$ come from the different parts of the CHY expression. The individual factors will be dealt with in appendix \ref{fact app}. We will only be interested in the dominant terms which will be the ones with the lowest power of $s$. Below is a summary of the powers of $s$ for the $(DF)^2$ theory contrasted with ordinary Yang-Mills:

\begin{center}

\begin{tabular}{c|c|c}
&$(DF)^2$ &  Yang-Mills  \label{side med tabel}   \\
\hline 
 \hline 
   $\frac{d^n\sigma}{\mathrm{vol[SL}(2,\mathbb{C})]}$                      & $s^{n_L-n_R-3}$   &  $s^{n_L-n_R-3}$\\ \hline 
 $\prod_i\phantom{}'\delta\left(\sum_{j\neq i}\frac{p_i\cdot p_j}{\sigma_{ij}}\right)$               &    $s^{n_L-n_R-2}$   &    $s^{n_L-n_R-2}$ \\  \hline 
 $\frac{\mathrm{Tr}\left(T^{a_{1}}T^{a_{2}}\cdots T^{a_{n}}\right)}{\sigma_{12}\sigma_{23}\cdots \sigma_{n1}}$                &          $s^{-n_L+n_R+2}$        &  $s^{-n_L+n_R+2}$\\ \hline
$\mathrm{Pf}'M_n$ & --- & $s^{-n_L+n_R+2}$ \\ \hline
$W_{\underbrace{11\cdots1}_n}$ & $s^{-n_L+n_R}$ & --- \\ 
\hline 
 \hline 
Total & $s^{-3}$ & $s^{-1}$
\end{tabular}

\end{center}

We see that the $(DF)^2$ theory has an extra factor of $s^{-2}$ compared to Yang-Mills, which is to be expected since this theory has double poles while Yang-Mills only has single poles. In the $q_R^2\to0$ limit, the amplitude of the $(DF)^2$ theory then become proportional to

\begin{align}
\int ds \frac{1}{s^3}\delta\left(\sum_{i\in R}\frac{v_n-v_i}{v_n}\sum_{j\in L}\frac{2p_i\cdot p_j}{u_jv_i}s^2-q_R^2\right)=&{}\frac{\sum_{i\in R}\frac{v_n-v_i}{v_n}\sum_{j\in L}\frac{p_i\cdot p_j}{u_jv_i}}{q_R^4}.\label{ds}
\end{align}

The numerator can be understood as the product of the $w_{(i)}$-functions for the new on-shell state. We therefore introduce polarization vectors for the intermediate state that has gone on-shell:

\begin{align}
\epsilon_{q_L}\cdot q_L&=0,\\
\epsilon_{q_R}\cdot q_R&=0,\\
\sum_{+/-}\epsilon_{q_L}^\mu\epsilon_{q_R}^\nu&=-2\eta^{\mu\nu}+\cdots.
\end{align}

Here $\cdots$ indicate terms proportional to $q_L^\mu$ or $q_R^\nu$. These terms vanish as each lower point amplitude is gauge-invariant. Equation \eqref{ds} can then be written as:

\begin{align}
\int ds\frac{1}{s^3}\delta\left(\sum_{i\in R}\frac{v_n-v_i}{v_n}\sum_{j\in L}\frac{2p_i\cdot p_j}{u_jv_i}s^2-q_R^2\right)=&{}\sum_{+/-}\frac{\sum_{i\in R}\frac{\epsilon_{q_R}\cdot p_i(v_n-v_i)}{v_iv_n}\sum_{j\in L}\frac{\epsilon_{q_L}\cdot p_j}{u_j}}{q_R^4}\label{ds2}
\end{align}

The numerator is equivalent to two $w_{(i)}$-functions with the $u$ and $v$ variables corresponding to the new on-shell state both having been fixed to 0.

The remaining details can be found in appendix \ref{fact app}. Putting them all together, one arrives at the conclusion that \eqref{DF^2 CHY} does indeed satisfy the correct factorization properties:

\begin{align}
\mathcal{A}_n^{ (DF)^2}(L,R)\bigg|_{q_R^2\to0}=&{}\sum_{+/-}\mathcal{A}_{n_L}^{ (DF)^2}(L,q_L^{a_L})\frac{-i\delta^{a_{q_L}a_{q_R}}}{q_R^4}\mathcal{A}_{n-n_L}^{ (DF)^2}(q_R^{a_R},R)\label{amp fact}
\end{align}

As a final comment about factorization, let us focus on some terms that do not play a role in \eqref{amp fact}, but are nonetheless interesting. They are some of the sub-leading terms from the color part of the CHY formula. The only terms that contribute to \eqref{amp fact} are those where the color generators in the trace separate nicely into one product of generators for the $L$ set and one product of generators for the $R$ set. As a shorthand, we could denote these as the $\mathrm{Tr}(LR)$-terms. However, one could also consider the $\mathrm{Tr}(LRLR)$-terms. Such terms do not generate a pole in Yang-Mills theory as they correspond to having an intermediate state which is not in the adjoint representation of the gauge group. However they do generate a simple pole in the $(DF)^2$ theory, which is to be expected since this theory does in fact contain particles that are not in the adjoint representation, the scalars.

To conclude, this section showed that \eqref{DF^2 CHY} factorizes into two amplitudes of the same form when a $1/q^4$ propagator was put on-shell. It also showed that the expression for the amplitude requires that the theory contain particles that are in a different representation of the gauge group than the adjoint. Both these observations support the claim that \eqref{DF^2 CHY} is in fact the correct $n$-point amplitude for the $(DF)^2$ theory.

% These are poles coming from the scalars in the theory, which as we mentioned in the beginning we would not focus on. The reason for the usual color factors accompanying these poles is the unusual color structure of the scalars in the theory.

% These terms correspond to a scalar propagator going on-shell. Since they only give rise to a $q^{-2}$-pole, their poles become sub-leading compared to the gluonic poles and one would expect not to see them. However because of the unusual color structure of the scalars, they generate poles that cannot be generate by the gluons, and hence these poles are possible to disentangle from the rest. 

\subsection{Conformal gravity amplitudes}

Conformal gravity can be found through combining the $(DF)^2$ theory described in section \ref{DFtheory} with standard super Yang-Mills in the KLT relations \cite{Kawai:1985xq}. In the CHY formalism, one can simply replace the color factor in \eqref{DF^2 CHY} with the reduced Pfaffian from Yang-Mills:

\begin{align}
\mathcal{A}_n^{\rm CG}=&{}\int \frac{d^n\sigma}{\mathrm{vol[SL}(2,\mathbb{C})]}\prod_i\phantom{}'\delta\left(\sum_{j\neq i}\frac{p_i\cdot p_j}{\sigma_{ij}}\right)W_{\underbrace{11\cdots1}_n}\mathrm{Pf}'M_n\label{CHY CG}
\end{align}

This should be the $D$-dimensional formula for conformal gravity (up to some overall constant). As a simple check for this formula let us point out that using the factorization properties of the reduced Pfaffian:

\begin{align}
\mathrm{Pf}'M_n\sim s^{n_R-n_L+2}\left(\prod_{i=1}^{n_L}u_i^2\right)\mathrm{Pf}'M_L\mathrm{Pf}'M_R,
\end{align}
\noindent it is straightforward to show that the formula factorizes correctly.% It is also easy to check for amplitudes with few external legs in four dimensions, this formula matches the expression found by 

Another simple check is to focus on the 4-dimensional MHV amplitudes. It is believed that in this case, there is only one relevant solution to the scattering equations	\cite{Monteiro:2013rya,Weinzierl:2014vwa,Naculich:2014naa}. It can be written in terms of spinors as follows:

\begin{align}
\sigma_i=&{}\frac{\langle i1\rangle\langle2\chi\rangle}{\langle i\chi\rangle\langle21\rangle}.
\end{align}

Here $|\chi\rangle$ is an arbitrary spinor not collinear with $|1\rangle$ or $|2\rangle$. This solution was proven to give the correct $n$-point MHV amplitude for Yang-Mills theory and Einstein gravity in \cite{Du:2016blz} where it was also shown that, at least up to 9-point, the other solutions to the scattering equations make the reduced Pfaffian vanish. Compared to those two theories, the only new element in \eqref{CHY CG} is the function $W_{11\cdots1}$ which, on this particular solution to the scattering equations and assuming that particles 1 and 2 are the only negative helicity gluons, can be written as:

\begin{align}
W_{\underbrace{11\cdots1}_n}=&{}\left(\frac{\langle21\rangle}{\langle1\chi\rangle\langle2\chi\rangle}\right)^n\left(\prod_{i=1}^n\langle i\chi\rangle^2\right)\left(\prod_{j=3}^n\sum_{k\neq j}\frac{[jk]\langle k\eta\rangle^2}{\langle jk\rangle\langle j\eta\rangle^2}\right),\label{W MHV}
\end{align}
\noindent where $|\eta\rangle$ is another arbitrary spinor not necessarily identical to $|\chi\rangle$. %Together with results from the other papers mentioned previously, it is then straigthforward to show in the MHV case \eqref{CHY CG} matches the expression found by Berkovits and Witten \cite{Berkovits:2004jj}:

By combining \eqref{W MHV} with results from the previously mentioned papers, one easily arrives at the following results for the scattering of gravitons among themselves and scattering between gravitons and scalars:

\begin{align}
\mathcal{A}_n^{\rm CG}(1^-2^-3^+\cdots n^+)=&{}\langle12\rangle^4\left(\prod_{j=3}^n\sum_{k\neq j}\frac{[jk]\langle k\eta\rangle^2}{\langle jk\rangle\langle j\eta\rangle^2}\right),\nonumber\\
\mathcal{A}_n^{\rm CG}(1^-2^\varphi 3^+\cdots n^+)=&{}\langle12\rangle^4\left(\prod_{j=2}^n\sum_{k\neq j}\frac{[jk]\langle k\eta\rangle^2}{\langle jk\rangle\langle j\eta\rangle^2}\right),\\
\mathcal{A}_n^{\rm CG}(1^\varphi 2^\varphi 3^+\cdots n^+)=&{}\langle12\rangle^4\left(\prod_{j=1}^n\sum_{k\neq j}\frac{[jk]\langle k\eta\rangle^2}{\langle jk\rangle\langle j\eta\rangle^2}\right)\nonumber
\end{align}

These amplitudes exactly match the expression found by Berkovits and Witten \cite{Berkovits:2004jj}, and thus we see that \eqref{CHY CG} can in fact be seen as a $D$-dimensional generalization of the Berkovits-Witten formula.

As a side note let us point out that this way of simplifying the CHY formulation in 4-dimensional MHV case will not work for the $(DF)^2$ theory. This is due to the fact that the function $W_{11\cdots1}$ is just a product of functions for each individual on-shell leg (the $w_{(i)}$'s from equation \eqref{Psi one index}), which means that for the function to be zero, one of these functions will have to be zero. These functions only depend on the helicity of the given external leg and not on all the other helicities. So if we imagine that a given solution to the scattering equations does not contribute to the all plus amplitudes because it sets $w_1$ to 0, then all other amplitudes where the helicity of particle 1 is positive will also not get contributions from this solution to the scattering equations.

We should also note that of supersymmetrizing \eqref{CHY CG} is essentially the same as the problem for Yang-Mills theory since the supersymmetry in the $R^2$ theory derives from this theory (see equation \eqref{KLT CSG}). If it is possible to construct a simple CHY-formulation for the amplitudes of $\mathcal{N}=4$ super Yang-Mills, it should therefore be straightforward to construct a supersymmetric version of equation \eqref{CHY CG} as well.

\section{Ambitwistor Interpretation}\label{ambitwist sec}

The fact that the
 amplitudes of conformal gravity and the $(DF)^2$ theory can be written as CHY formulae suggests that there should be ambitwistor string theories \cite{Mason:2013sva,Casali:2015vta} corresponding to them.
In this section we will briefly review ambitwistor string theory and show which specific choices of the worldsheet action lead to the amplitudes given in equations \eqref{DF^2 CHY} and \eqref{CHY CG}.

\subsection{Review}

The ambitwistor string theories  can be thought of as chiral worldsheet models describing the interactions of massless states. In the simplest example, bosonic strings, the action is given by
\begin{equation}
S_\mathrm{B} = \frac{1}{2\pi}\int \mathrm{d}^2 \sigma \left( P_\mu \bar\partial X^\mu - \frac{1}{2}eP^2\right),
\label{bosambi}
\end{equation}
where $X^\mu$  ($\mu = 0$ to $D-1$) denotes the string coordinates in the $D$-dimensional target space, $P_\mu$ are their conjugate momenta and $e$ is a Lagrange multiplier enforcing the constraint $P^2=0$.

Because of this first-class constraint, the model is invariant under the following local symmetry, in addition to reparameterization invariance:
\begin{equation}
\delta X^\mu = \alpha P^\mu\,, \qquad \delta P_\mu = 0\,, \qquad \delta e = \bar\partial \alpha\,,
\end{equation}
for some transformation parameter $\alpha$. One can use this symmetry to gauge-fix $e = 0$, and then the standard BRST procedure yields the gauge-fixed action
\begin{equation}
S_\mathrm{B}^\star = \frac{1}{2\pi}\int \mathrm{d}^2 \sigma \left( P_\mu \bar\partial X^\mu + b\bar\partial c + \widetilde{b}\bar\partial\widetilde{c}\right),
\end{equation}
together with the BRST charge
\begin{equation}
Q = \frac{1}{2\pi\mathrm{i}}\oint \mathrm{d}\sigma \left(cT - bc\partial c +\frac{1}{2}\widetilde{c}P^2\right),
\label{BRSTbos}
\end{equation}
where $T$ is the complete energy-momentum tensor (matter $+$ ghosts), $(b,c)$ are the usual (anti)ghosts of string theory and $(\widetilde{b},\widetilde{c})$ are the (anti)ghosts corresponding to the extra gauge symmetry.

Physical states correspond to vertex operators in the cohomology of $Q$, which in this case contains only\footnote{In this paper we consider only plane-wave states, even though higher-derivative theories typically contain other types of states such as those of the form $A\cdot X\, \mathrm{e}^{\mathrm{i} p\cdot X}$.}
\begin{equation}
V = c \widetilde{c}\,P_\mu P_\nu \epsilon^{\mu\nu} \mathrm{e}^{\mathrm{i} p\cdot X}
\label{Vbos}
\end{equation}
and its integrated version
\begin{equation}
U = \int \mathrm{d}^2 \sigma\,\bar\delta(p\cdot P) \,P_\mu P_\nu \epsilon^{\mu\nu} \mathrm{e}^{\mathrm{i} p\cdot X}\,.
\label{Ubos}
\end{equation}
BRST-closedness requires $p^2 = p_\mu\epsilon^{\mu\nu}=0$, while the analysis of BRST-exact states implies the gauge transformation $\delta \epsilon^{\mu\nu} = p^{(\mu}\epsilon^{\nu)}$ for some parameter $\epsilon^\mu$ such that $p_\mu\epsilon^\mu = 0$. Thus, these operators correspond to an on-shell graviton.

However, if one computes the correlation function containing three unintegrated vertex operators, the result does not agree with the expected three-point amplitude coming from Einstein gravity. In fact, it is of order six in the momenta. In \cite{Mason:2013sva}, the authors could not interpret the result in terms of any known theory of gravity, although they mention that it could be related to a (Weyl)${}^3$ vertex. The tree-level $n$-point function is given by
\begin{equation}
\mathcal{A}_n=\int \frac{d^n\sigma}{\mathrm{vol[SL}(2,\mathbb{C})]}{\prod_i}^\prime\bar\delta\left( p^{(i)}\cdot P(\sigma_i)\right) \prod_{j=1}^n \epsilon_{(j)}^{\mu\nu}P_\mu(\sigma_j)P_\nu(\sigma_j)\,,
\end{equation}
with $P_\mu$ constrained to take its value as $P_\mu(\sigma) = \sum_{i=1}^n p^{(i)}_\mu/(\sigma-\sigma_i)$. Note that, using the language introduced in section 3, this amplitude can be cast as
\begin{equation}
\mathcal{A}_n=\int \frac{d^n\sigma}{\mathrm{vol[SL}(2,\mathbb{C})]}{\prod_i}^\prime\bar\delta\left( p^{(i)}\cdot P(\sigma_i)\right) {\left(W_{\underbrace{11\cdots1}_n}\right)}^2,
\end{equation}
and the appearance of the $W_{11\cdots1}$ function squared indicates that this theory will be the result of squaring the $(DF)^2$ theory via the double copy.

 This purely bosonic model can be generalized in many different ways. To do so, the standard procedure consists of adding two other terms to the action  (\ref{bosambi}), $S_\mathrm{L}$ and $S_\mathrm{R}$, which ultimately correspond to the left and right integrands in CHY formulae (cf. (\ref{CHY generel})).

In perhaps the most successful example, both $S_\mathrm{L}$ and $S_\mathrm{R}$ are RNS-like fermion systems, with the important difference that in the ambitwistor case all worldsheet fields are left-moving (holomorphic). The complete action is given by:
\begin{equation}
S_\mathrm{B} + S_\mathrm{L} + S_\mathrm{R} = \frac{1}{2\pi}\int \mathrm{d}^2 \sigma \left( P_\mu \bar\partial X^\mu - \frac{1}{2}eP^2  +\frac{1}{2}\sum_{r=1,2} (\Psi_{r\mu} \bar\partial \Psi^\mu_r - 2\chi_r P_\mu\Psi_r^\mu)\right),
\label{actiontypeII}
\end{equation}
where $\Psi_1^\mu, \Psi_2^\mu$ are the worldsheet fermions and $\chi_1, \chi_2$ are fermionic Lagrange multipliers for the fermionic constraints $P\cdot\Psi_1, P\cdot\Psi_2$.

Gauge-fixing the Lagrange multipliers to zero via the BRST procedure, one ends up with the usual RNS-like bosonic (anti)ghosts $(\beta_1,\gamma_1)$ and $(\beta_2,\gamma_2)$, in addition to the same (anti)ghosts as before. The BRST charge is now given by
\begin{equation}
Q_{(\Psi_1, \Psi_2)} = \frac{1}{2\pi\mathrm{i}}\oint \mathrm{d}\sigma\left(cT - bc\partial c +\frac{1}{2}\widetilde{c}P^2 +\sum_r (\gamma_r\Psi_r\cdot P +\widetilde{b}\gamma_r\gamma_r)\right)
\label{BRSTtypeII}
\end{equation}
and its cohomology contains the vertex operator
\begin{equation}
V^{(-1)}_{(\Psi_1, \Psi_2)} = c\widetilde{c} \,\mathrm{e}^{\mathrm{i}p\cdot X} \prod_r \delta(\gamma_r) \Psi_r\cdot \epsilon_r\,,
\label{vertextypeII}
\end{equation}
together with corresponding  picture-number-zero or integrated versions, where $\epsilon_1^\mu, \epsilon_2^\nu$ combine to form the graviton, Kalb--Ramond and dilaton polarizations.
One can show that the tree-level $n$-point correlation function of these vertex operators gives rise to the CHY formula (\ref{Einstein Gravity}) when restricted to gravitons.

Another possibility for $(S_\mathrm{L}, S_\mathrm{R})$ is to replace one of the fermionic systems of the previous model with an action for a generic current algebra, $S_\mathrm{C}$. Then one can define the currents $J_I$ satisfying the OPE
\begin{equation}
J_I(\sigma_1) J_J(\sigma_2) \sim \frac{\ell}{(\sigma_1-\sigma_2)^2}\delta_{IJ} + \frac{1}{\sigma_1-\sigma_2}f_{IJ}{}^K J_K(\sigma_2)\,,
\end{equation}
where $\ell$ is the so-called level of the algebra and $f_{IJ}{}^K$ are the structure constants of the gauge group. The BRST charge of this model has the same form as (\ref{BRSTtypeII}), with the obvious differences that now the sum over $r$ comprises only one term and the energy-momentum tensor is the one corresponding to the new gauge-fixed action.

This theory is reminiscent of the usual heterotic string theory, and its spectrum also contains two sectors: the gauge one and the gravity one. However, the latter does not correspond to the usual Neveu--Schwarz sector of heterotic strings, and in particular it contains a 3-form potential whose interpretation was unclear in the original work by Mason and Skinner. 
 In the gauge sector, the following vertex operator belongs to the cohomology of $Q$:
\begin{equation}
V^{(-1)}_{(\Psi, J)} = c\widetilde{c} \,\delta(\gamma) \Psi\cdot \epsilon\,J_I T^I  \mathrm{e}^{\mathrm{i}p\cdot X}  \,,
\label{V_Psi,J}
\end{equation}
where $T^I$ denotes the generators of the gauge group. BRST invariance imposes $p^2=p\cdot\epsilon=0$, and the vertex operator is BRST-trivial if $\epsilon_\mu \propto p_\mu$. Therefore, it describes an on-shell gluon.

When restricted to single-trace contributions, the tree-level $n$-point correlation function involving (\ref{V_Psi,J}) (and the other versions of this vertex operator, as appropriate) is equal to the CHY formula (\ref{YM CHY}) for gluons.

\subsection{\boldmath{$(S_\mathrm{L}, 0)$-models}}\label{s_l,0 afsnit}

From the review in the previous subsection, it should be clear that there is a correspondence between the choice of $(S_\mathrm{L}, S_\mathrm{R})$, the vertex operators and the correlation functions of a given ambitwistor string. We summarize the results presented so far in the following table.

\begin{table}[h!]
\centering
\begin{tabular}{c|c|c}
$S_\mathrm{L/R}$ &  Vertex & $I_\mathrm{L/R}$    \\
\hline 
 \hline 
   $0$                      &  $\epsilon \cdot P$   &   $W_{\underbrace{11\cdots1}_n}$\\ \hline 
 $\Psi$               &    $\delta(\gamma) \epsilon\cdot\Psi$   &    Pf${}^\prime(M_n)$ \\  \hline 
 $J$                &          $T^I J_I$        &  color factor
\end{tabular}
\end{table}

\noindent In the above, $0$ signifies that $S_\mathrm{L}$ or $S_\mathrm{R}$ are absent from the model, e.g. $(0,0)$ represents the bosonic ambitwistor string. Moreover, ``Vertex'' denotes the contribution to the simplest vertex operator and $I_\mathrm{L/R}$ the two different parts of the integrand in the CHY formulation of amplitudes (cf. (\ref{CHY generel})).
 More precisely, the (single-trace) tree-level $n$-point correlation function of any $(S_\mathrm{L}, S_\mathrm{R})$-model gives rise to a CHY formula containing $I_\mathrm{L}$ and $I_\mathrm{R}$.

%Given the previous exposition, it becomes evident that the amplitudes in the $(DF)^2$ theory can be described by the ambitwistor model $(J,0)$.

Thus, by comparing with \eqref{DF^2 CHY}, we see that the CHY formula for the $(DF)^2$-theory can be obtained via the ambitwistor model $(J,0)$, while a comparison with \eqref{CHY CG} leads to the conclusion that the CHY formula for conformal supergravity can be obtained through the  model $(\Psi,0)$.

Since, to the best of our knowledge, models of the type $(S_\mathrm{L},0)$ have not yet been explored in the literature,  it is worth to discuss them in a bit more detail. In the $(J,0)$ case, the action is given by
\begin{equation}
S_{(J,0)} = \frac{1}{2\pi}\int \mathrm{d}^2 \sigma \left( P_\mu \bar\partial X^\mu - \frac{1}{2}eP^2 + \mathcal{L}_\mathrm{C}\right),
\label{S_J,0}
\end{equation}
where $\mathcal{L}_\mathrm{C}$ is the Lagrangian corresponding to a generic current algebra. The gauge-fixing procedure is almost identical to the one for the bosonic case, and we are left with the BRST-charge
\begin{equation}
Q_{(J,0)} = \frac{1}{2\pi\mathrm{i}}\oint \mathrm{d}\sigma\left(cT - bc\partial c +\frac{1}{2}\widetilde{c}P^2\right),
\label{BRSTJ,0}
\end{equation}
which looks exactly the same as (\ref{BRSTbos}), but now $T$ includes the energy-momentum tensor $T_\mathrm{C}$ corresponding to $\mathcal{L}_\mathrm{C}$. Accordingly, the central charge receives a contribution $c_\mathrm{C}$ from the gauge sector, and is given by $c_{(J,0)} = 2(D-26) + c_\mathrm{C}$. Thus, one can  make $c_{(J,0)}$ vanish in a given number of dimensions by choosing the current algebra appropriately. However, we need not concern ourselves much about this since we only work at tree level.

The cohomology of $Q_{(J,0)}$ contains the vertex operator
\begin{equation}
V_{(J,0)} = c\widetilde{c} \,P \cdot \epsilon\,  \mathrm{e}^{\mathrm{i}p\cdot X}  J_I T^I \,,
\label{V_J,0}
\end{equation}
together with its integrated version --- which as usual amounts to replacing the ghosts with $\int\mathrm{d}^2 \sigma\,\bar\delta(p\cdot P)$. This expression is BRST-invariant if and only if $p^2=p\cdot\epsilon=0$, and $\epsilon^\mu \propto p^\mu$ renders it BRST-trivial, hence it corresponds to an on-shell gluon. It is easy to see that the tree-level $n$-point correlation function computed with these operators gives rise to \eqref{DF^2 CHY}.

Note that the cohomology also contains gravity states, a feature common to all known ambitwistor string theories. In this case, the graviton vertex operators are identical to the ones in the bosonic model, given in (\ref{Vbos}) and (\ref{Ubos}), and thus the 3-point amplitude exhibits the same (Weyl)${}^3$ behavior. As anticipated in the introduction, it is a general property of $(S_\mathrm{L},0)$-models that the states and tree-level amplitudes obtainable from one such model can also be obtained from an $(S_\mathrm{L},J)$-model,
and the appearance of gravity states in the $(0,J)$-model is just a consequence of that. 
By the same token, the $(J,0)$-model  can be identified with a sector of the more general $(J,\tilde{J})$-model, which contains bi-adjoint scalars transforming under two potentially different gauge groups. It is remarkable that the ambitwistor framework allows  such a truncation, i.e. that some sectors can be treated as theories on their own.
We will encounter another example of that in the following.

\vspace{1cm}

Let us now discuss the $(\Psi,0)$ ambitwistor string, which gives rise to the tree-level $n$-point amplitude in \eqref{CHY CG}.
The action of the model is given by
\begin{equation}
S_{(\Psi,0)} = \frac{1}{2\pi}\int \mathrm{d}^2 \sigma \left( P_\mu \bar\partial X^\mu - \frac{1}{2}eP^2  +\frac{1}{2} \Psi_{\mu} \bar\partial \Psi^\mu - \chi P_\mu\Psi^\mu\right).
\end{equation}
After gauge-fixing $e=\chi=0$, one gets the BRST charge
\begin{equation}
Q_{(\Psi,0)} = \frac{1}{2\pi\mathrm{i}}\oint \mathrm{d}\sigma\left(cT - bc\partial c +\frac{1}{2}\widetilde{c}P^2 +\gamma \Psi^\mu  P_\mu +\widetilde{b}\gamma^2\right),
\label{BRSTPsi,0}
\end{equation}
whose cohomology contains the vertex operator
\begin{equation}
V^{(-1)}_{(\Psi, 0)} = c\widetilde{c} \, \delta(\gamma)  \epsilon_1^\mu \epsilon_2^\nu \Psi_\mu P_\nu  \mathrm{e}^{\mathrm{i}p\cdot X}\,,
\end{equation}
together with corresponding  picture-number-zero or integrated versions, where $\epsilon_1^\mu, \epsilon_2^\nu$ combine to form the graviton, Kalb--Ramond and dilaton polarizations.
Restricting to gravitons, one can show that
the tree-level $n$-point correlation function of these vertex operators gives rise to the CHY formula \eqref{CHY CG}. However, since the central charge is computed to give $c_{(\Psi, 0)} = \frac{5}{2}D-41$, it is not possible to make sense of this model beyond tree level, in any (integer) number of  dimensions.

Note that, at tree level, this model is equivalent to the gravity sector of the heterotic ambitwistor string, given by $(\Psi,J)$. Indeed, the current-algebra part of the heterotic model is inert in the gravity sector, which implies that the cohomology and correlation functions are the same as those in the $(\Psi,0)$ model. In particular, the $(\Psi,0)$  model also contains the unexpected (from the Einstein-gravity point of view) massless 3-form first encountered in \cite{Mason:2013sva}, whose picture-number $-1$ vertex operator is given by
\begin{equation}
V^{(-1)}_{\mathrm{3-form}}=c\widetilde{c}\,\delta(\gamma) A_{\mu\nu\rho} \Psi^\mu \Psi^\nu \Psi^\rho \mathrm{e}^{\mathrm{i}p\cdot X}\,,
\end{equation}
with $p^\mu A_{\mu\nu\rho}=0$.
Therefore, we conclude that the gravity sector of the heterotic ambitwistor string describes conformal supergravity, and  it is then natural to interpret that theory as a generalization of Witten's twistor string theory. We will come back to this point shortly.

\vspace{1cm}

Finally, we would like to discuss the more exotic case of the $((\Psi_1,\Psi_2),0)$ ambitwistor string. This is reminiscent of the $(\Psi_1,\Psi_2)$ model, and indeed the action and BRST operator are the same as (\ref{actiontypeII}) and (\ref{BRSTtypeII}), respectively. Hence, one would naively think that the spectrum and correlation functions of the two models are identical.

However, putting both fermion systems on the same side of the model translates into having weaker GSO-like conditions. To make this point clearer, consider the following state:
\begin{equation}
V^{(-1)}_{((\Psi_1,\Psi_2),\,0)} = c\widetilde{c} \, \delta(\gamma_1)  \delta(\gamma_2)\, \epsilon\cdot P\, \mathrm{e}^{\mathrm{i}p\cdot X}\,.
\label{newstate}
\end{equation}
For $p^2=p\cdot\epsilon=0$, this state is BRST-invariant, and $\epsilon^\mu \propto p^\mu$ renders it trivial, as usual. In the $(\Psi_1,\Psi_2)$ model, this state is projected out of the physical spectrum, since in that case one requires physical states to have an even number of $\{\gamma_1, \Psi_1\}$ and an even number of $\{\gamma_2, \Psi_2\}$ --- cf. (\ref{vertextypeII}), for example. In the $((\Psi_1,\Psi_2),0)$ model, the GSO-like projection requires that the number of $\{\gamma_1,\gamma_2,\Psi_1,\Psi_2\}$ be even, and thus both (\ref{vertextypeII}) and (\ref{newstate})  are considered physical. Since there is no current algebra in this particular model, the state in \eqref{newstate} corresponds to a U(1)-field, i.e. a photon.

One can show that the  tree-level $n$-point correlation function of these photon states gives
\begin{equation}
\mathcal{A}_n=\int \frac{d^n\sigma}{\mathrm{vol[SL}(2,\mathbb{C})]}{\prod_i}^\prime\bar\delta\left( p^{(i)}\cdot P(\sigma_i)\right) \left. W_{\underbrace{11\cdots1}_n}\right. {\left(\mathrm{Pf}{}^\prime(M_{A,n})\right)}^2,\label{DF^2 photon amplitude}
\end{equation}
where $M_{A,n}$ is an $n$ by $n$ matrix identical to one of the submatrices of the bigger matrix $M_n$ defined in \eqref{SubMatricesM_n}.
From this discussion, it is evident that one more row can be added to the table above \cite{Casali:2015vta}:
%\begin{table}[h!]
%\centering
\begin{center}
\begin{tabular}{c|c|c}
$S_\mathrm{L/R}$ &  Vertex & $I_\mathrm{L/R}$    \\
\hline 
 \hline & \\
   $(\Psi_1,\Psi_2)$             &  $\delta(\gamma_1)  \delta(\gamma_2)$            &  ${\left(\mathrm{Pf}{}^\prime(M_{A,n})\right)}^2$
\end{tabular}
\end{center}
%\end{table}

Let us now consider the amplitude in \eqref{DF^2 photon amplitude} from the quantum field theory point of view. It arises from combining the $(DF)^2$ theory with the non-linear sigma model in the KLT relations.\footnote{The non-linear sigma model corresponds to the $((\Psi_1,\Psi_2),J)$  ambitwistor string, as can be seen from the table displayed in the introduction.} By inspecting the amplitude, we find that up to four points the simplest Lagrangian for this theory is given by:

\begin{align}
\tfrac{1}{\sqrt{-g}}\mathcal{L}=&{}\frac{1}{2\kappa^2}R+\tfrac{1}{4}\left(\nabla_\mu F_{\nu\rho}\right)\left(\nabla^\mu F^{\nu\rho}\right)+\tfrac{1}{8}RF_{\mu\nu}F^{\mu\nu}-\tfrac{1}{6}\kappa^2\left(\nabla_\mu F_{\sigma\lambda}\right)\left(\nabla^\sigma F^{\mu\nu}\right)F_{\nu\rho}F^{\rho\lambda}\nonumber\\
&+\tfrac{1}{48}\kappa^2\left(\nabla_\mu F_{\nu\rho}\right)\left(\nabla^\mu F^{\nu\rho}\right)F_{\sigma\lambda}F^{\sigma\lambda}+\cdots.
\end{align}

We will refer to this theory as the $(DF)^2$-photon theory. Note that the ordinary Einstein gravity appears as part of this Lagrangian and that the coupling constant for its self-interaction is the same as for its interaction with the gravitons. From the ambitwistor string theory point of view, the appearance of Einstein gravity is fairly obvious since both the vertices \eqref{vertextypeII} and \eqref{newstate} are allowed in the $((\Psi_1,\Psi_2),0)$ model. From the quantum field theory perspective, it is less clear how the product of the $(DF)^2$ theory and the non-linear sigma model can give rise to a spin-2 field. Nonetheless, the $(DF)^2$-photon is bound to interact with regular Einstein gravity, as can be seen by the following factorization argument.

Consider an amplitude of $2n$ $(DF)^2$-photons, group the photons into $n$ pairs and take the limit where the propagator for each pair goes on-shell. In this scenario, the amplitude in \eqref{DF^2 photon amplitude} behaves in the following way:

\begin{align}
\Bigg(\prod_{i\in\{1,3,\cdots 2n-1\}}\lim_{p_i\cdot p_{i+1}\to 0}p_i\cdot p_{i+1}\Bigg)\mathcal{A}^{(DF)^2-photon}_{2n}\propto&{}
% \Bigg(\prod_{i\in\{1,3,\cdots 2n-1\}}\varepsilon_i\cdot p_{i+1}\varepsilon_{i+1}\cdot p_{i}\Bigg)\\
%&
\int \frac{d^n\sigma}{\mathrm{vol[SL}(2,\mathbb{C})]}{\prod_i}^\prime\bar\delta\left(\sum_{j\neq i}\frac{p_i\cdot p_j}{\sigma_{ij}}\right)  \mathrm{det}{}^\prime(\widetilde{M}_{A,2n})\nonumber,
\end{align}

\noindent where the matrix $\widetilde{M}_{A,2n}$ can be written in the following form (where $i$ and $j$ only run over the odd numbers):

\begin{align}
\widetilde{M}_{A,2n}^{i,j}&=\left\{\begin{array}{cc}\frac{(p_i+p_{i+1})\cdot (p_j+p_{j+1})}{\sigma_{ij}}&\text{for }i\neq j\\
0&\text{for }i=j\end{array}\right.,&\widetilde{M}_{A,2n}^{i+n,j+n}&=\left\{\begin{array}{cc}\frac{p_i\cdot p_{j+1}}{\sigma_{ij}}&\text{for }i\neq j\\
0&\text{for }i=j\end{array}\right.,\\
\widetilde{M}_{A,2n}^{i+n,j}&=\left\{\begin{array}{cc}\frac{p_i\cdot (p_j+p_{j+1})}{\sigma_{ij}}&\text{for }i\neq j\\
-\sum_{j\neq i}\frac{p_i\cdot (p_j+p_{j+1})}{\sigma_{ij}}&\text{for }i=j\end{array}\right.,&
\widetilde{M}_{A,2n}^{i+n,j}&=\left\{\begin{array}{cc}\frac{(p_i+p_{i+1})\cdot p_{j+1}}{\sigma_{ij}}&\text{for }i\neq j\\
-\sum_{i\neq j}\frac{(p_i+p_{i+1})\cdot p_{j+1}}{\sigma_{ij}}&\text{for }i=j\end{array}\right..\nonumber
\end{align}

By comparing with the formula for Einstein gravity \eqref{Einstein Gravity}, one sees that this is the amplitude of $n$ gravitons with momenta $p_i+p_{i+1}$ where the polarization vectors have been replaced by $p_i^{(\mu}p_{i+1}^{\nu)}$. This makes it clear also from the quantum field theory perspective that the $(DF)^2$ photon couples to Einstein gravity.

%With these three theories in hand, we can expand the usual matrix of theories arising from ambitwistor strings by adding a new row/column as in figure \ref{teorimatrix}.

\subsection{Connection to Witten's twistor string}

Even though we only discuss bosonic states in this paper, it should be said that the spectrum of the $(\Psi, J)$ ambitwistor string theory also contains fermions and is in fact supersymmetric  --- see  \cite{me&renann2} for a description in the pure-spinor context. In ten dimensions, the gauge sector corresponds to SYM, while the gravity sector must be equivalent to the $R^2$ conformal supergravity studied by de Roo in \cite{deRoo:1991at} --- see also \cite{Bergshoeff:1982az} ---, since the action presented in that paper is supposed to be unique.

From our point of view, it is then natural to interpret this theory as a $D$-dimensional generalization of Witten's twistor string theory \cite{Witten:2003nn}. In four dimensions, the gauge sector describes $\mathcal{N}=4$ SYM, while the gravity sector reduces to the conformal supergravity sector analyzed by Berkovits and Witten \cite{Berkovits:2004jj}. Indeed, the CHY formula \eqref{CHY CG} can be obtained from the gravity sector of this ambitwistor theory.
Note also that a massless 3-form has no propagating degrees of freedom in four dimensions. In summary, we have the following table of approaches to the same theory:

\begin{table}[h!]
\centering
\begin{tabular}{c|c|c}
Double-copy &  Ambitwistor &  in $D=4$    \\
\hline 
 \hline 
   (DF)${}^2$ $\otimes$ SYM  &  $(\Psi,0)$                      & Berkovits--Witten sector\\ \hline 
 ((DF)${}^2$ $+$ $\phi^3$) $\otimes$ SYM  &  $(\Psi,J)$     &  Witten's twistor string
\end{tabular}
\end{table}
\noindent where $\phi^3$ stands for the bi-adjoint scalar theory, whose amplitudes can be obtained in the CHY representation through the $(J,\tilde{J})$ ambitwistor string. It would be very interesting to obtain a more direct relation between the heterotic ambitwistor string and the twistor string studied by Berkovits and Witten, for example at the level of vertex operators. We plan to address this question in future work.

\section{Conclusions}

In this paper, we introduced three new, elegant CHY-type formulae and provided an ambi\-twistor string interpretation for each of them. The string actions are all of the type $(S_L,0)$ so, together with the bosonic ambitwistor string, they form an entire new row/column in the matrix of possible ambitwistor models.

First we considered the $(DF)^2$ theory introduced in \cite{conformal}. The CHY formulation of this theory is simple and exposes a property of the amplitudes that is far from obvious from the Feynman diagram perspective, namely the absence of $\varepsilon_i \cdot \varepsilon_j$ terms.

The second theory we considered was an $R^2$ theory of gravity which in $D=4$ becomes conformal gravity. Our work can therefore be seen as a $D$-dimensional generalization of the paper \cite{Berkovits:2004jj} by Berkovits and Witten, and our CHY formulation of the amplitudes does in fact reduce to their result in the appropriate limit.

Finally, we looked at the $(DF)^2$-photon theory. This theory arose naturally from our studies of the previous two theories. An interesting feature of this theory is that the photon couples to regular Einstein gravity. This may seem surprising since the theory can be described using the KLT relations as the product of the non-linear sigma model and  the previously mentioned $(DF)^2$ theory. Simplistically one would expect to get at most spin-1 fields running around in such a theory since the non-linear sigma model contains only scalars and the $(DF)^2$ theory consists of scalars and gluons. This expectation is however wrong and, as demonstrated in section \ref{s_l,0 afsnit}, one can in fact get an $n$-point Einstein gravity amplitude from the appropriate limit of an amplitude of $2n$ $(DF)^2$-photons. It will be interesting to study this theory further and try to understand this in detail. Central to this surprising fact are certainly the scalars in the $(DF)^2$ theory and their unusual color structure.

The role of the scalars is in general interesting, if somewhat mysterious. They are essential for the $(DF)^2$ theory to satisfy the color-kinematics duality, but their strange color structure leads to non-planar diagrams making contributions to tree-level amplitudes. For instance this means that in the four-point amplitudes, the numerator $n_s$ could get a term proportional $1/u$ (terms like this can of course be removed through redefinitions of the numerators, but only in exchange for similarly weird terms in the other numerators). This in turn makes the interpretation of the function of the fields in the double copy a bit hazy, because it means that an internal graviton carrying momentum $p_1+p_2$ somehow is the product of a gluon with the same momentum and a scalar carrying momentum $p_1+p_3$. Perhaps a closer look at the amplitudes of the scalars will provide some answers. It should be fairly straightforward to get some of the amplitudes from the ${\rm Tr}(LRLR)$-terms arising in the factorization limit as described towards the end of section \ref{fact sub}.

\subsection*{Acknowledgments}
We are grateful to Henrik Johansson for suggesting the problem, sharing details about his work with Josh Nohle and for providing comments on the draft. TA acknowledges financial support from
the Knut and Alice Wallenberg Foundation under grant 2015.0083. OTE is supported by the Knut and Alice Wallenberg Foundation under grant KAW~2013.0235.

%Putting \eqref{measure factorization}, \eqref{fix delta functions}, \eqref{R delta}, \eqref{L delta}, \eqref{color factorization}, \eqref{Psi R} and \eqref{Psi L} together, we get:

%\begin{align}
%\mathcal{A}_n^{ (DF)^2}\bigg|_{q_R^2\to0}=&{}4ig^n\int(-1)^{n_L}\left(\prod_{i=3}^{n_L}du_i\right)\left(\prod_{i=n_L+1}^{n-2}dv_i\right)ds\,s^{n_L-n_R-3}\label{measure factorization}\\
%&\frac{2(u_1-u_2)(v_{n-1}-v_n)(-s^4+u_1u_2v_{n-1}v_n)}{v_{n-1}\prod_{i=1}^{n_L}u_i^2},
%\end{align}

%The content of the delta functions will depend on whether the corresponding external particle is in the group $R$ or the group $L$. For particles in the group $R$, the

%The factorization of \eqref{} can easily be found by following the steps taken in \cite{Cachazo:2013iea} with some 

\appendix

\section{Factorization Details}\label{fact app}

In this appendix we will give some of the details to the factorization calculation done in section \ref{fact sub}. We begin with the delta functions:

\begin{align}
\prod_i\phantom{}'\delta\left(\sum_{j\neq i}\frac{p_i\cdot p_j}{\sigma_{ij}}\right)\equiv&{}\sigma_{kl}\sigma_{lm}\sigma_{mk}\prod_{i\neq k,l,m}\delta\left(\sum_{j\neq i}\frac{p_i\cdot p_j}{\sigma_{ij}}\right).
\end{align}

As mentioned in section \ref{scat eq sec}, the prime indicates that three delta functions have been removed and replaced by a product of differences between the $\sigma$'s corresponding to the removed delta functions. We choose to remove the delta functions corresponding to the three fixed $\sigma$'s (though strictly speaking one could have made a different choice). The product of differences then becomes:

\begin{align}
(\sigma_1-\sigma_2)(\sigma_2-\sigma_n)(\sigma_n-\sigma_1)=&{}\frac{(u_1-u_2)(s^2-v_nu_2)(s^2-v_nu_1)}{u_1^2u_2^2s}.\label{fix delta functions}
\end{align}

Now we turn to the delta functions. Notice that one of the delta functions for the particles in the $R$ is used to impose the behaviour of $s$ $i.e.$ it becomes the delta function in equation \eqref{ds}. To see how this comes about, just consider the scattering equations for the particles in the $R$ set:

\begin{align}
\sum_{j=1}^n\frac{p_i\cdot p_j}{\sigma_i-\sigma_j}
=&{}s\sum_{j\in R}\frac{p_i\cdot p_j}{v_i-v_j}+s\frac{p_i\cdot q_R}{v_i}+s^3\sum_{j\in L}\frac{p_i\cdot p_j}{v_iu_j\left(v_i-\frac{s^2}{u_j}\right)}\nonumber.
\end{align}

If we multiply this $v_i(v_n-v_i)/sv_n$ and sum over all of the particles belonging to $R$, this becomes:

\begin{align}
\sum_{i\in R}\sum_{j=1}^n\frac{p_i\cdot p_j}{\sigma_{ij}}
=&{}-\tfrac{1}{2}q_R^2+s^2\sum_{i\in R}\frac{v_n-v_i}{v_n}\sum_{j\in L}\frac{p_i\cdot p_j}{u_j\left(v_i-\frac{s^2}{u_j}\right)}.\nonumber
\end{align}

This is what imposes the behaviour of $s$ in \eqref{s^2 behave}. In total the delta functions for the $R$ set becomes:

\begin{align}
\prod_{i\in R/\{n\}}\delta\left(\sum_{k=1}^n\frac{p_i\cdot p_k}{\sigma_{ij}}\right)=&{}\frac{2v_{n-1}(v_n-v_{n-1})}{v_n}s^{1-n_R}\delta\left(\sum_{i\in R}\frac{v_n-v_i}{v_n}\sum_{j\in L}\frac{2p_i\cdot p_j}{u_jv_i}s^2-q_R^2\right)\label{R delta}\\
&\prod_{i\in R/\{n-1,n\}}\delta\left(\sum_{j\in R}\frac{p_i\cdot p_j}{v_i-v_j}+\frac{2p_i\cdot q_R}{v_i}+\mathcal{O}(s^2)\right),\nonumber
\end{align}

The delta functions for the particles in the $L$ set has a straightforward under the shift and become:

\begin{align}
\prod_{i\in L/\{1,2\}}\delta\left(\sum_{j=1}^n\frac{p_i\cdot p_j}{\sigma_{ij}}\right)=&{}s^{n_L-2}\prod_{i\in L/\{1,2\}}\delta\left(-u_i^2\left(\sum_{j\in L}\frac{p_i\cdot p_j}{u_i-u_j}+\frac{p_i\cdot q_L}{u_i}\right)+\mathcal{O}(s^2)\right)\label{L delta}
\end{align}

Putting the factors of $s$ together from \eqref{fix delta functions}, \eqref{R delta} and \eqref{L delta}, we get that the dominant behaviour will be $s^{n_L-n_R-2}$ as in the table on page \pageref{side med tabel}.

The integration measure behaves as follows under the change of variables:

\begin{align}
\frac{1}{\mathrm{vol[SL}(2,\mathbb{C})]}\prod_{i=1}^nd\sigma_i=&{}(-1)^{n_L}\left(\prod_{i=3}^{n_L}du_i\right)\left(\prod_{j=n_L+1}^{n-2}dv_i\right)ds\,s^{n_L-n_R-3}\label{measure factorization}\\
&\frac{2(u_1-u_2)(v_{n-1}-v_n)(-s^4+u_1u_2v_{n-1}v_n)}{v_{n-1}\prod_{i=1}^{n_L}u_i^2},\nonumber
\end{align}

Let us now proceed to the color part of the formula:

\begin{align*}
&\sum_{\beta\in S_n/\sigma_n}\frac{\mathrm{Tr}\left(T^{a_{\beta(1)}}T^{a_{\beta(2)}}\cdots T^{a_{\beta(n)}}\right)}{\sigma_{\beta(1)\beta(2)}\sigma_{\beta(2)\beta(3)}\cdots \sigma_{\beta(n)\beta(1)}}.
\end{align*}

The different terms in the sum depend differently upon $s$ so we will begin by determining which have the lowest power of $s$. Each term contains $n$ factors of $(\sigma_i-\sigma_j)^{-1}$. If both $i$ and $j$ belong to $L$, such a factor will contribute with $s^{-1}$ while if they both belongs to $R$, it will contribute with $s$. If $i$ belongs to $L$ and $j$ belongs $R$ or vice versa, such a factor will contribute with $s$. As a consequence the terms with as few factors of $(\sigma_i-\sigma_j)^{-1}$ where $i$ and $j$ belongs to different sets, will be the terms will the lowest power in $s$. This is perhaps not surprising from the point of view of the color factor as the amplitude is thus split into a product of two planar amplitudes with one only containing the particles from the set $L$ plus an intermediate state and the other only the ones from the set $R$ plus the intermediate state.

\begin{align}
&\sum_{\beta\in S_n/\sigma_n}\frac{\mathrm{Tr}\left(T^{a_{\beta(1)}}T^{a_{\beta(2)}}\cdots T^{a_{\beta(n)}}\right)}{\sigma_{\beta(1)\beta(2)}\sigma_{\beta(2)\beta(3)}\cdots \sigma_{\beta(n)\beta(1)}}\nonumber\\
=&{}(-1)^{n_L}s^{n_R-n_L+2}\left(\prod_{i=1}^{n_L}u_i^2\right)
\sum_{\alpha\in S_{n_L}}\frac{\mathrm{Tr}(T^{a_{\alpha(1)}}\cdots T^{a_{\alpha(n_L)}}T^{a_{q_L}})}{u_{\alpha(1)}u_{\alpha(1),\alpha(2)}\cdots u_{\alpha(n_L-1),\alpha(n_L)}u_{\alpha(n_L)}}\label{color factorization}\\
&\delta^{a_{q_L}a_{q_R}}\sum_{\beta\in S_{n_R}}\frac{\mathrm{Tr}(T^{a_{q_R}}T^{a_{\beta(n_L+1)}}\cdots T^{a_{\beta(n)}})}{v_{\beta(n_L+1)}v_{\beta(n_L+1),\beta(n_L+2)}\cdots v_{\beta(n-1),\beta(n)}v_{\beta(n)}}+\mathcal{O}(s^{n_R-n_L+4}).\nonumber
\end{align}

The factor involving the traces over the gauge group generators look exactly as one would expect if we imagine the $u$ and $v$ variables corresponding to the new on-shell state to have been fixed to zero. We note that the dominant term is proportional to $s^{n_R-n_L+2}$ as mentioned in the table on page \pageref{side med tabel}.

Finally, we consider the function $W_{11\cdots1}$ or rather the individual functions that it is a product of, the $w_i$'s. If $i$ belongs to the set $R$, this function becomes:

\begin{align}
w_i=&{}s\sum_{j\in R,j\neq i}\frac{\epsilon_i\cdot p_j(v_j-v_r)}{(v_r-v_i)(v_i-v_j)}-s\frac{\epsilon_i\cdot q_Rv_r}{(v_r-v_i)v_i}+\mathcal{O}(s^3),\label{Psi R}
\end{align}
\noindent while it becomes the following for $i$ belonging to $L$:

\begin{align}
w_i=&{}-\frac{u_i^2}{s}\sum_{j\in L,j\neq i}\frac{\epsilon_i\cdot p_j(u_j-u_r)}{(u_r-u_i)(u_i-u_j)}+\frac{u_i^2}{s}\frac{\epsilon_i\cdot q_Lu_r}{(u_r-u_i)u_i}+\mathcal{O}(s).\label{Psi L}
\end{align}

We see that it both cases the dominant terms depend only on the other particles in the same set in addition to a term depending on the momentum of the internal propagator that has gone on-shell. From the above expressions we see that $W_{11\cdots1}$ will contribute with a factor of $s^{n_R-n_L}$ as mentioned in the table on page \pageref{side med tabel}.


\begin{thebibliography}{99}



%\cite{Cachazo:2013hca}
%Oprindelig CHY
\bibitem{Cachazo:2013hca}
  F.~Cachazo, S.~He and E.~Y.~Yuan,
  %``Scattering of Massless Particles in Arbitrary Dimensions,''
  Phys.\ Rev.\ Lett.\  {\bf 113} (2014) no.17,  171601
  %doi:10.1103/PhysRevLett.113.171601
  [arXiv:1307.2199 [hep-th]].
  %%CITATION = doi:10.1103/PhysRevLett.113.171601;%%
  %180 citations counted in INSPIRE as of 19 Jan 2017


%\cite{Cachazo:2013iea}
%den du bruger til faktorisering
\bibitem{Cachazo:2013iea}
  F.~Cachazo, S.~He and E.~Y.~Yuan,
  %``Scattering of Massless Particles: Scalars, Gluons and Gravitons,''
  JHEP {\bf 1407} (2014) 033
  %doi:10.1007/JHEP07(2014)033
  [arXiv:1309.0885 [hep-th]].
  %%CITATION = doi:10.1007/JHEP07(2014)033;%%
  %158 citations counted in INSPIRE as of 19 Jan 2017


%\cite{Cachazo:2014xea}
\bibitem{Cachazo:2014xea}
  F.~Cachazo, S.~He and E.~Y.~Yuan,
  %``Scattering Equations and Matrices: From Einstein To Yang-Mills, DBI and NLSM,''
  JHEP {\bf 1507} (2015) 149
  doi:10.1007/JHEP07(2015)149
  [arXiv:1412.3479 [hep-th]].
  %%CITATION = doi:10.1007/JHEP07(2015)149;%%
  %98 citations counted in INSPIRE as of 24 Feb 2017

%\cite{Mason:2013sva}
\bibitem{Mason:2013sva}
  L.~Mason and D.~Skinner,
  %``Ambitwistor strings and the scattering equations,''
  JHEP {\bf 1407} (2014) 048
  doi:10.1007/JHEP07(2014)048
  [arXiv:1311.2564 [hep-th]].
  %%CITATION = doi:10.1007/JHEP07(2014)048;%%
  %106 citations counted in INSPIRE as of 23 Feb 2017
	
	%\cite{Casali:2015vta}
\bibitem{Casali:2015vta}
  E.~Casali, Y.~Geyer, L.~Mason, R.~Monteiro and K.~A.~Roehrig,
  %``New Ambitwistor String Theories,''
  JHEP {\bf 1511} (2015) 038
  doi:10.1007/JHEP11(2015)038
  [arXiv:1506.08771 [hep-th]].
  %%CITATION = doi:10.1007/JHEP11(2015)038;%%
  %31 citations counted in INSPIRE as of 27 Feb 2017


\bibitem{conformal} H.~Johansson and J.~Nohle,
  %``Conformal Gravity from Gauge Theory,''
  arXiv:1707.02965 [hep-th].


\bibitem{Fradkin:1985am} 
  E.~S.~Fradkin and A.~A.~Tseytlin,
  %``Conformal Supergravity,''
  Phys.\ Rept.\  {\bf 119}, 233 (1985).
  doi:10.1016/0370-1573(85)90138-3









%\cite{Kawai:1985xq}
\bibitem{Kawai:1985xq}
  H.~Kawai, D.~C.~Lewellen and S.~H.~H.~Tye,
  %``A Relation Between Tree Amplitudes of Closed and Open Strings,''
  Nucl.\ Phys.\ B {\bf 269} (1986) 1.
  doi:10.1016/0550-3213(86)90362-7
  %%CITATION = doi:10.1016/0550-3213(86)90362-7;%%
  %557 citations counted in INSPIRE as of 20 Feb 2017


%\cite{Berkovits:2004jj}
\bibitem{Berkovits:2004jj}
  N.~Berkovits and E.~Witten,
  %``Conformal supergravity in twistor-string theory,''
  JHEP {\bf 0408} (2004) 009
  doi:10.1088/1126-6708/2004/08/009
  [hep-th/0406051].
  %%CITATION = doi:10.1088/1126-6708/2004/08/009;%%
  %177 citations counted in INSPIRE as of 22 Feb 2017

%\cite{Witten:2003nn}
\bibitem{Witten:2003nn} 
  E.~Witten,
  %``Perturbative gauge theory as a string theory in twistor space,''
  Commun.\ Math.\ Phys.\  {\bf 252}, 189 (2004)
  doi:10.1007/s00220-004-1187-3
  [hep-th/0312171].
  %%CITATION = doi:10.1007/s00220-004-1187-3;%%
  %932 citations counted in INSPIRE as of 25 Apr 2017


\bibitem{Maldacena:2011mk} 
  J.~Maldacena,
  %``Einstein Gravity from Conformal Gravity,''
  arXiv:1105.5632 [hep-th].




\bibitem{Carrasco:2013ypa} 
  J.~J.~M.~Carrasco, R.~Kallosh, R.~Roiban and A.~A.~Tseytlin,
  %``On the U(1) duality anomaly and the S-matrix of N=4 supergravity,''
  JHEP {\bf 1307}, 029 (2013)
  doi:10.1007/JHEP07(2013)029
  [arXiv:1303.6219 [hep-th]].











%\cite{Simmons:1989zs}
\bibitem{Simmons:1989zs}
  E.~H.~Simmons,
  %``Dimension-six Gluon Operators as Probes of New Physics,''
  Phys.\ Lett.\ B {\bf 226} (1989) 132.
  doi:10.1016/0370-2693(89)90301-8
  %%CITATION = doi:10.1016/0370-2693(89)90301-8;%%
  %28 citations counted in INSPIRE as of 29 Aug 2017

%\cite{Simmons:1990dh}
\bibitem{Simmons:1990dh}
  E.~H.~Simmons,
  %``Higher dimension gluon operators and hadronic scattering,''
  Phys.\ Lett.\ B {\bf 246} (1990) 471.
  doi:10.1016/0370-2693(90)90632-G
  %%CITATION = doi:10.1016/0370-2693(90)90632-G;%%
  %23 citations counted in INSPIRE as of 29 Aug 2017
	
	%\cite{Cho:1993eu}
\bibitem{Cho:1993eu}
  P.~L.~Cho and E.~H.~Simmons,
  %``Looking for gluon substructure at the tevatron,''
  Phys.\ Lett.\ B {\bf 323} (1994) 401
  doi:10.1016/0370-2693(94)91238-6
  [hep-ph/9307345].
  %%CITATION = doi:10.1016/0370-2693(94)91238-6;%%
  %27 citations counted in INSPIRE as of 29 Aug 2017
	
	%\cite{Duff:1991ad}
\bibitem{Duff:1991ad}
  A.~Duff and D.~Zeppenfeld,
  %``Probing QCD via four jet decays of the Z boson,''
  Z.\ Phys.\ C {\bf 53} (1992) 529.
  doi:10.1007/BF01625915
  %%CITATION = doi:10.1007/BF01625915;%%
  %11 citations counted in INSPIRE as of 29 Aug 2017
	
	%\cite{Dreiner:1991xi}
\bibitem{Dreiner:1991xi}
  H.~K.~Dreiner, A.~Duff and D.~Zeppenfeld,
  %``How well do we know the three gluon vertex?,''
  Phys.\ Lett.\ B {\bf 282} (1992) 441.
  doi:10.1016/0370-2693(92)90666-R
  %%CITATION = doi:10.1016/0370-2693(92)90666-R;%%
  %18 citations counted in INSPIRE as of 29 Aug 2017

%\cite{Dixon:1993xd}
\bibitem{Dixon:1993xd}
  L.~J.~Dixon and Y.~Shadmi,
  %``Testing gluon selfinteractions in three jet events at hadron colliders,''
  Nucl.\ Phys.\ B {\bf 423} (1994) 3
   Erratum: [Nucl.\ Phys.\ B {\bf 452} (1995) 724]
  doi:10.1016/0550-3213(94)90563-0, 10.1016/0550-3213(95)00450-7
  [hep-ph/9312363].
  %%CITATION = doi:10.1016/0550-3213(94)90563-0, 10.1016/0550-3213(95)00450-7;%%
  %24 citations counted in INSPIRE as of 29 Aug 2017

%\cite{Barreiro:2012aw}
\bibitem{Barreiro:2012aw}
  L.~A.~Barreiro and R.~Medina,
  %``Revisiting the S-matrix approach to the open superstring low energy effective lagrangian,''
  JHEP {\bf 1210} (2012) 108
  doi:10.1007/JHEP10(2012)108
  [arXiv:1208.6066 [hep-th]].
  %%CITATION = doi:10.1007/JHEP10(2012)108;%%
  %24 citations counted in INSPIRE as of 29 Aug 2017
	
	%\cite{Barreiro:2013dpa}
\bibitem{Barreiro:2013dpa}
  L.~A.~Barreiro and R.~Medina,
  %``RNS derivation of N-point disk amplitudes from the revisited S-matrix approach,''
  Nucl.\ Phys.\ B {\bf 886} (2014) 870
  doi:10.1016/j.nuclphysb.2014.07.015
  [arXiv:1310.5942 [hep-th]].
  %%CITATION = doi:10.1016/j.nuclphysb.2014.07.015;%%
  %31 citations counted in INSPIRE as of 29 Aug 2017
	
	%\cite{Boels:2016xhc}
\bibitem{Boels:2016xhc}
  R.~H.~Boels and R.~Medina,
  %``Graviton and gluon scattering from first principles,''
  Phys.\ Rev.\ Lett.\  {\bf 118} (2017) no.6,  061602
  doi:10.1103/PhysRevLett.118.061602
  [arXiv:1607.08246 [hep-th]].
  %%CITATION = doi:10.1103/PhysRevLett.118.061602;%%
  %9 citations counted in INSPIRE as of 29 Aug 2017




%\cite{Kleiss:1988ne}
\bibitem{Kleiss:1988ne}
  R.~Kleiss and H.~Kuijf,
  %``Multi - Gluon Cross-sections and Five Jet Production at Hadron Colliders,''
  Nucl.\ Phys.\ B {\bf 312} (1989) 616.
  doi:10.1016/0550-3213(89)90574-9
  %%CITATION = doi:10.1016/0550-3213(89)90574-9;%%
  %231 citations counted in INSPIRE as of 22 Aug 2017

%\cite{DelDuca:1999iql}
\bibitem{DelDuca:1999iql}
  V.~Del Duca, A.~Frizzo and F.~Maltoni,
  %``Factorization of tree QCD amplitudes in the high-energy limit and in the collinear limit,''
  Nucl.\ Phys.\ B {\bf 568} (2000) 211
  doi:10.1016/S0550-3213(99)00657-4
  [hep-ph/9909464].
  %%CITATION = doi:10.1016/S0550-3213(99)00657-4;%%
  %149 citations counted in INSPIRE as of 22 Aug 2017
	
	%\cite{DelDuca:1999rs}
\bibitem{DelDuca:1999rs}
  V.~Del Duca, L.~J.~Dixon and F.~Maltoni,
  %``New color decompositions for gauge amplitudes at tree and loop level,''
  Nucl.\ Phys.\ B {\bf 571} (2000) 51
  doi:10.1016/S0550-3213(99)00809-3
  [hep-ph/9910563].
  %%CITATION = doi:10.1016/S0550-3213(99)00809-3;%%
  %195 citations counted in INSPIRE as of 22 Aug 2017


%\cite{Bern:2008qj}
\bibitem{Bern:2008qj}
  Z.~Bern, J.~J.~M.~Carrasco and H.~Johansson,
  %``New Relations for Gauge-Theory Amplitudes,''
  Phys.\ Rev.\ D {\bf 78} (2008) 085011
  doi:10.1103/PhysRevD.78.085011
  [arXiv:0805.3993 [hep-ph]].
  %%CITATION = doi:10.1103/PhysRevD.78.085011;%%
  %415 citations counted in INSPIRE as of 27 Feb 2017


%\cite{Sondergaard:2009za}
\bibitem{Sondergaard:2009za}
  T.~Sondergaard,
  %``New Relations for Gauge-Theory Amplitudes with Matter,''
  Nucl.\ Phys.\ B {\bf 821} (2009) 417
  doi:10.1016/j.nuclphysb.2009.07.002
  [arXiv:0903.5453 [hep-th]].
  %%CITATION = doi:10.1016/j.nuclphysb.2009.07.002;%%
  %39 citations counted in INSPIRE as of 22 Aug 2017
	
%\cite{BjerrumBohr:2009rd}
\bibitem{BjerrumBohr:2009rd}
  N.~E.~J.~Bjerrum-Bohr, P.~H.~Damgaard and P.~Vanhove,
  %``Minimal Basis for Gauge Theory Amplitudes,''
  Phys.\ Rev.\ Lett.\  {\bf 103} (2009) 161602
  doi:10.1103/PhysRevLett.103.161602
  [arXiv:0907.1425 [hep-th]].
  %%CITATION = doi:10.1103/PhysRevLett.103.161602;%%
  %207 citations counted in INSPIRE as of 22 Aug 2017
	
	
	%\cite{Feng:2010my}
\bibitem{Feng:2010my}
  B.~Feng, R.~Huang and Y.~Jia,
  %``Gauge Amplitude Identities by On-shell Recursion Relation in S-matrix Program,''
  Phys.\ Lett.\ B {\bf 695} (2011) 350
  doi:10.1016/j.physletb.2010.11.011
  [arXiv:1004.3417 [hep-th]].
  %%CITATION = doi:10.1016/j.physletb.2010.11.011;%%
  %128 citations counted in INSPIRE as of 22 Aug 2017
	
	%\cite{Britto:2004ap}
\bibitem{Britto:2004ap}
  R.~Britto, F.~Cachazo and B.~Feng,
  %``New recursion relations for tree amplitudes of gluons,''
  Nucl.\ Phys.\ B {\bf 715} (2005) 499
  doi:10.1016/j.nuclphysb.2005.02.030
  [hep-th/0412308].
  %%CITATION = doi:10.1016/j.nuclphysb.2005.02.030;%%
  %706 citations counted in INSPIRE as of 22 Aug 2017

%\cite{Britto:2005fq}
\bibitem{Britto:2005fq}
  R.~Britto, F.~Cachazo, B.~Feng and E.~Witten,
  %``Direct proof of tree-level recursion relation in Yang-Mills theory,''
  Phys.\ Rev.\ Lett.\  {\bf 94} (2005) 181602
  doi:10.1103/PhysRevLett.94.181602
  [hep-th/0501052].
  %%CITATION = doi:10.1103/PhysRevLett.94.181602;%%
  %873 citations counted in INSPIRE as of 22 Aug 2017
	
	
%\cite{Lam:2016tlk}
\bibitem{Lam:2016tlk}
  C.~S.~Lam and Y.~P.~Yao,
  %``Evaluation of the Cacha\sigmao-He-Yuan gauge amplitude,''
  Phys.\ Rev.\ D {\bf 93} (2016) no.10,  105008
  %doi:10.1103/PhysRevD.93.105008
  [arXiv:1602.06419 [hep-th]].
  %%CITATION = doi:10.1103/PhysRevD.93.105008;%%
  %14 citations counted in INSPIRE as of 18 Jan 2017

%\cite{He:2016iqi}
\bibitem{He:2016iqi}
  S.~He and Y.~Zhang,
  %``New Formulas for Amplitudes from Higher-Dimensional Operators,''
  arXiv:1608.08448 [hep-th].
  %%CITATION = ARXIV:1608.08448;%%
  %3 citations counted in INSPIRE as of 18 Jan 2017

%\cite{Baadsgaard:2015voa}
\bibitem{Baadsgaard:2015voa}
  C.~Baadsgaard, N.~E.~J.~Bjerrum-Bohr, J.~L.~Bourjaily and P.~H.~Damgaard,
  %``Integration Rules for Scattering Equations,''
  JHEP {\bf 1509} (2015) 129
  doi:10.1007/JHEP09(2015)129
  [arXiv:1506.06137 [hep-th]].
  %%CITATION = doi:10.1007/JHEP09(2015)129;%%
  %32 citations counted in INSPIRE as of 23 Feb 2017

	%\cite{Bjerrum-Bohr:2016juj}
\bibitem{Bjerrum-Bohr:2016juj}
  N.~E.~J.~Bjerrum-Bohr, J.~L.~Bourjaily, P.~H.~Damgaard and B.~Feng,
  %``Analytic representations of Yang–Mills amplitudes,''
  Nucl.\ Phys.\ B {\bf 913} (2016) 964
  doi:10.1016/j.nuclphysb.2016.10.012
  [arXiv:1605.06501 [hep-th]].
  %%CITATION = doi:10.1016/j.nuclphysb.2016.10.012;%%
  %13 citations counted in INSPIRE as of 23 Feb 2017

%\cite{Zhou:2017mfj}
\bibitem{Zhou:2017mfj}
  K.~Zhou, J.~Rao and B.~Feng,
  %``Derivation of Feynman Rules for Higher Order Poles Using Cross-ratio Identities in CHY Construction,''
  JHEP {\bf 1706} (2017) 091
  doi:10.1007/JHEP06(2017)091
  [arXiv:1705.04783 [hep-th]].
  %%CITATION = doi:10.1007/JHEP06(2017)091;%%
  %1 citations counted in INSPIRE as of 28 Aug 2017

%\cite{Monteiro:2013rya}
\bibitem{Monteiro:2013rya}
  R.~Monteiro and D.~O'Connell,
  %``The Kinematic Algebras from the Scattering Equations,''
  JHEP {\bf 1403} (2014) 110
  doi:10.1007/JHEP03(2014)110
  [arXiv:1311.1151 [hep-th]].
  %%CITATION = doi:10.1007/JHEP03(2014)110;%%
  %49 citations counted in INSPIRE as of 02 Mar 2017
	
	%\cite{Weinzierl:2014vwa}
\bibitem{Weinzierl:2014vwa}
  S.~Weinzierl,
  %``On the solutions of the scattering equations,''
  JHEP {\bf 1404} (2014) 092
  doi:10.1007/JHEP04(2014)092
  [arXiv:1402.2516 [hep-th]].
  %%CITATION = doi:10.1007/JHEP04(2014)092;%%
  %33 citations counted in INSPIRE as of 02 Mar 2017

	%\cite{Naculich:2014naa}
\bibitem{Naculich:2014naa}
  S.~G.~Naculich,
  %``Scattering equations and BCJ relations for gauge and gravitational amplitudes with massive scalar particles,''
  JHEP {\bf 1409} (2014) 029
  doi:10.1007/JHEP09(2014)029
  [arXiv:1407.7836 [hep-th]].
  %%CITATION = doi:10.1007/JHEP09(2014)029;%%
  %56 citations counted in INSPIRE as of 02 Mar 2017
	
	%\cite{Du:2016blz}
\bibitem{Du:2016blz}
  Y.~j.~Du, F.~Teng and Y.~s.~Wu,
  %``CHY formula and MHV amplitudes,''
  JHEP {\bf 1605} (2016) 086
  doi:10.1007/JHEP05(2016)086
  [arXiv:1603.08158 [hep-th]].
  %%CITATION = doi:10.1007/JHEP05(2016)086;%%
  %10 citations counted in INSPIRE as of 02 Mar 2017






\bibitem{me&renann2} T.~Azevedo and R.~L.~Jusinskas,
  %``Connecting the ambitwistor and the sectorized heterotic strings,''
  arXiv:1707.08840 [hep-th].


\bibitem{deRoo:1991at} 
  M.~de Roo,
  %``The R**2 action in d = 10 conformal supergravity,''
  Nucl.\ Phys.\ B {\bf 372}, 243 (1992).
  doi:10.1016/0550-3213(92)90319-7

\bibitem{Bergshoeff:1982az} 
  E.~Bergshoeff, M.~de Roo and B.~de Wit,
  %``Conformal Supergravity in Ten-dimensions,''
  Nucl.\ Phys.\ B {\bf 217}, 489 (1983).
  doi:10.1016/0550-3213(83)90159-1








\end{thebibliography}
\end{document}